\documentclass[twocolumn]{aastex6}
\usepackage{amsmath}
\usepackage{soul}
\usepackage{bm}
\newcommand{\mdot}{\dot{\mathcal{M}}}

\shorttitle{}
\shortauthors{}

\begin{document}

\title{Effects of Turbulence on the Critical Conditions of Explosion}
\title{How Turbulence Enables Core-Collapse Supernova Explosions}

\author{Quintin A. Mabanta\altaffilmark{1}}
\author{Jeremiah W. Murphy\altaffilmark{2}}

\altaffiltext{1}{Florida State University, qam13b@my.fsu.edu}
\altaffiltext{2}{Florida State University, jwmurphy@fsu.edu}

\begin{abstract}
An important result in core-collapse supernova (CCSN) theory
  is that spherically-symmetric,
one-dimensional simulations routinely fail to explode, yet multi-dimensional simulations
  often explode. Numerical
investigations suggest that turbulence
eases the condition for explosion, but how is not fully
  understood. We develop a turbulence model for neutrino-driven
  convection, and show that this turbulence model reduces
  the condition for explosions by about 30\%, in concordance with multi-dimensional simulations.
 In addition, we identify which turbulent
  terms enable explosions.  Contrary to
    prior suggestions, turbulent ram pressure is not the dominant factor in reducing the condition for explosion.
Instead, there are many contributing factors,
  ram pressure being only one of them, but the dominant factor is
turbulent dissipation (TD).   Primarily, TD provides extra heating,
adding significant thermal pressure, and reducing the condition for explosion.  The source of
this TD power is turbulent kinetic energy,
which ultimately derives its energy from the higher
potential of an unstable convective profile. Investigating a turbulence model in conjunction
  with an explosion condition enables insight that is difficult to
  glean from merely analyzing complex multi-dimensional simulations. An explosion
  condition presents a clear diagnostic to explain why stars explode, and the turbulence
  model allows us to explore how turbulence enables explosion. Though
  we find that turbulent dissipation is
a significant contributor to successful supernova explosions,
it is important to note that this work is to some extent qualitative.
Therefore, we suggest ways to further verify and validate our
  predictions with multi-dimensional simulations.
\end{abstract}


\section{Introduction}
\label{sec:introduction}

Understanding how massive stars end their lives still remains an
important astrophysical problem.  Observations indicate that most massive stars
likely explode as core-collapse supernovae \citep{li11,horiuchi11}, but the theoretical
details of the explosion mechanism are still uncertain.  The most
certain part of the theory is that the Fe core collapses, bounces at
nuclear densities forming a proto-neutron star, and launches a
shock wave \citep{janka16}.  However, this shock wave quickly stalls \citep{hillebrandt81,mazurek82,mck82}.  For over three decades,
the most prominent mechanism has been the delayed-neutrino mechanism
in which neutrinos reheat the matter below the stalled shock and
relaunch it into an explosive blast wave \citep{bethe85}.  For
all but the least massive stars, the neutrino mechanism fails
in one-dimensional, spherically-symmetric simulations
\citep{liebendorfer01a,liebendorfer01b,liebendorfer05b,rampp02,buras03,thompson03,kitaura06,buras06b,radice17} but multi-dimensional simulations do explode 
\citep{herant94,janka95,burrows95,janka96,burrows07b,melson15,dolence15,muller16,roberts16,bruenn16} and initial
analyses suggest that multi-dimensional instabilities and turbulence
aid the neutrino mechanism toward explosion \citep{murphy08b,marek09b,murphy11,hanke12,murphy13,radice16,couch15,melson15}.  Therefore, understanding
the explosion mechanism requires an understanding of the conditions
between failed and successful explosions and how turbulence aids the
explosion.  There appears to be a critical condition for explosion \citep{burrows93,murphy08b,murphy17},
and the critical condition for explosion is easier to obtain in multi-dimensional
simulations \citep{herant94,janka95,janka96,burrows95,murphy08b,melson15}.  In this manuscript, we propose an analytic model for turbulence and investigate how it reduces the condition for explosion. 

For a thorough review on the status and problems of CCSN theory, see \citet{muller17} and \citet{janka16}.  Here, we motivate our work with some of the most salient points.
We know that the core collapses, but not yet how the collapse reverses into explosion.
Whether a massive star explodes or not, the Fe core collapses
at the end of the massive star's life.
Prior to collapse, the overlying Si-burning layer adds Fe onto the iron core.  As the iron core grows in
size, the electrons which supply much of the electron degeneracy pressure
become more and more relativistic.  As the core nears the Chandrasekhar mass limit, the
relativistic electron degeneracy pressure becomes less effective at
supporting the core against gravitational collapse.  Meanwhile,
neutrino losses reduce the lepton number, decreasing the number of
electrons available to supply pressure.  The iron
core becomes gravitationally unstable and contracts down to a sea of nucleons, forming a proto-neutron
star, and is thus supported mostly by the strong force. At these nuclear densities, the equation of state for
  the core stiffens, and the core abruptly bounces, slamming into the
  rest of the star which is collapsing supersonically onto the
  bouncing proto-neutron star.  This creates
  an outward moving shock wave.   As the shock wave propagates outward, it loses
  energy via photodissociation of Fe, electron capture, and
neutrino losses. The shock stalls into an accretion shock, but
if the star is to explode, this stalled accretion shock must relaunch
into an explosive blast wave.

One proposed solution to relaunching the blast wave was the
delayed-neutrino mechanism \citep{bethe85}. During the stalled
  shock phase, the neutron star is cooling with a neutrino
  luminosity of a few $\times$ $10^{52}$ ergs/s, but only about 10\% of this
  luminosity is recaptured in a net heating region, the gain region. This volume is above the proto-neutron star but below the shock.  If neutrino heating were the only effect, then this would be sufficient to relaunch the explosion.  However, the region below the shock is
  in sonic contact, and so the structure satisfies a boundary-value
  problem.  While the neutrino heating adds heat that would drive
  explosion, the neutrino cooling at the base and the ram pressure of
  matter accreting through the shock keeps the shock stalled.  The
  hope has been that for high enough neutrino luminosities or low
  accretion rates, the neutrinos may overwhelm the ram pressure and
  relaunch the explosion.  Alas, spherically symmetric simulations show that this mechanism
fails in in all but the least massive progenitors \citep{liebendorfer01a,liebendorfer01b,liebendorfer05b,rampp02,buras03,thompson03,kitaura06,buras06b,muller17b,radice17}.  
 
While one-dimensional simulations fail, many multi-dimensional simulations
  seem to succeed (sometimes weakly) \citep{herant94,janka95,burrows95,janka96,burrows07b,melson15,dolence15,muller16,roberts16,bruenn16,burrows16}, and the best indications are that  turbulence plays an important role in aiding the delayed-neutrino
  mechanism toward explosion \citep{bethe85,janka96,marek09b,murphy13,murphy11}. Therefore, to truly understand the
  explosion mechanism of massive stars, we need to identify the
  conditions for explosion and how turbulence affects these
  conditions.

There are many attempts to characterize these conditions, some are
empirical \citep{oconnor11,ott13,ertl16}, some are
heuristic \citep{janka98,thompson00,thompson05,buras06,muller16b}, and others attempt
to derive a condition from fundamentals \citep{burrows93,pejcha12,murphy17}.  Of these, the most  illuminating to date has been the neutrino-luminosity and accretion-rate
  critical curve \citep{burrows93}.  During the stalled shock phase, \citet{burrows93} noted that
  one may derive steady-state solutions for the stalled shock
  structure.  The governing equations describe a boundary value
  problem in which the lower boundary is set by the neutron star
  surface (the neutrino-sphere) and the outer boundary is the shock.
  They parameterized the problem in terms of the accretion rate, $\dot{\mathcal{M}}$,
  onto the shock, and the neutrino luminosity, $L_\nu$, emanating from the
  core.  In this two-dimensional parameter space they found
  steady-state solutions below a critical curve.  Above this curve, they
did not find steady-state stalled solutions, and suggested, but did
not prove, that the solutions above the critical
neutrino-luminosity and accretion-rate curve are dynamic and explosive.


\citet{murphy17} take a step closer to proving that the solutions above the curve
are explosive by showing that the only steady solutions above this curve indeed have a positive shock velocity ($v_s > 0$).  \citet{burrows93} focused on just $\dot{\mathcal{M}}$ and $L_\nu$, but there are
five parameters that define the steady solutions.  They are neutrino luminosity ($L_\nu$), mass accretion rate ($\dot{\mathcal{M}}$), neutron star mass ($M_{NS}$), neutron star radius ($R_{NS}$), and the neutrino temperature ($T_\nu$).
\citet{murphy17} point out that the critical condition is not a critical
curve but a critical hypersurface; most importantly, they find
that this critical hypersurface is described by one dimensionless
parameter, $\Psi$.  In essence, $\Psi$ is an integral condition related to
the balance of pressure and gravity behind the shock.  For a given set
of the five parameters, $\Psi$ may be negative, zero, or positive,
which correspond to $v_s < 0$, $v_s = 0$, and $v_s > 0$.  There is
always a minimum $\Psi$, and if $\Psi_{\rm min} < 0$, then there is
always a stable steady-state stalled solution such that $\Psi=0$.  The critical condition
is where $\Psi_{\rm min} = 0$.  Above this critical condition,
$\Psi_{\rm min} > 0$, and all steady solutions have $v_s > 0$.
Assuming that these $v_s > 0$ steady solutions correspond to
explosion, they use $\Psi_{\rm min}$ to define an explodability
parameter. 

Using one- and two-dimensional simulations \citet{murphy08b},
empirically confirmed that criticality is a useful condition for explosion in
core-collapse simulations.  Furthermore, they found that the critical condition is $\sim$30\%
lower in two-dimensional simulations.  Subsequently, others have confirmed these
findings and that the reduction is similar in three-dimensional simulations \citep{fernandez15,hanke12,dolence13,handy14}.

Initial indications are that turbulence causes this reduction but
these investigations were mostly suggestive and not
conclusive. Decades ago,
\citet{bethe90} recognized the potential importance of
neutrino-driven convection aiding explosion. This initial investigation suggested
that turbulent ram pressure behind the material would push against the
shock. In the early 90s, the first
two-dimensional simulations with crude neutrino
transport exploded while the one-dimensional
simulations did not. These investigators speculated that
neutrino-driven convection aided the explosion \citep{burrows95,janka96,janka01,colgate04}. 
  
\citet{blondin03} identified a new
instability that can also drive turbulence: the standing accretion
shock instability (SASI).  Linear analyses suggest
that this instability results from an advective-acoustic feedback
cycle \citep{fog00,fog06,sato09a,sato09b,guilet10}, and
subsequent investigations considered the possibility that the SASI
aids the delayed neutrino mechanism toward explosion \citep{marek09b,hanke12,hanke13,fernandez14}. Instead, \citet{murphy13} found that in simplified
simulations convection dominates just before and during neutrino-driven explosions. Further analyses with less
  simplistic neutrino approaches found that the SASI does dominate at times, but convection
likely dominates for most but not all explosion conditions 
\citep{muller16,radice16,burrows12,murphy13,murphy11}.
 
In one attempt to find the reason why turbulence aids explosion, \citet{murphy08b} found that entropy is higher in the
multi-dimensional case. At the time, they  as well as others suggested that a longer dwell time in the gain region causes this
higher entropy \citep{thompson05,marek09b,buras06}, however, these investigations did not show
  that these dwell-time distributions actually lead to enhanced
  entropy in the gain region. \citet{couch15}
revisited the idea of turbulent ram pressure being the main
multi-dimensional contribution, but speculated it as an
\textit{effective} ram pressure, not distinguishing between multi-dimensional effects. At this point many of these suggestions seem
  plausible, but it is not clear which, if any, explain why turbulence
aids explosions.  In fact, we will show that none of these
explanations truly captured the role of turbulence.  However, we do
note that the higher entropy profiles of multi-dimensional turbulence
should have hinted that turbulent dissipation plays an important role.

Part of the reason it was difficult to
assess how turbulence aids explosion is the complexity of
multi-dimensional simulations. In these large, non-linear simulations
it is difficult to isolate the causes and effects of turbulence. In
this paper, we propose a different, more illuminating approach.  We model turbulence in the context of the critical condition. Because we have direct control over how turbulence affects the equations, we can directly assess the causes and effects of our turbulence model in reducing the critical condition.

To do this, we extend the explodability parameter of \citet{murphy17} to multiple dimensions
  by including a turbulence model. Currently, a turbulence model exists for neutrino
  driven convection \citep{murphy11,murphy13}, but one does not yet exist for the SASI. Thus, the only viable avenue of an analytic investigation of turbulence is through neutrino-driven
convection. Investigations on how turbulence driven by the
  SASI affects the condition will have to wait until we have a valid
  turbulence model for the SASI. Therefore, we use the neutrino-driven convection turbulence model of \citet{murphy13} in the critical condition of \citet{murphy17}. 

In section~\ref{section2}, we revisit criticality, present the neutrino-driven
turbulence model, derive the equations, and describe the solution
method. Furthermore, we derive an analytic upper limit on the Reynolds
stress. In section~\ref{section3} we present how turbulence modifies the
structure of the post-shock flow and how it affects the critical
condition. Finally, in section~\ref{section4}, we conclude and discuss
implications for the core-collapse mechanism and future
investigations. Our main conclusions are that turbulent ram
  pressure is not the primary turbulent term aiding explosions, rather it is one of a few and the dominant
  turbulent effect aiding explosions is turbulent dissipation. 

\section{Methods}
\label{section2}



In this section, we outline the method for
  deriving the critical condition including the effects of
  neutrino-driven convection. Our primary goal is to incorporate a
  turbulence model in the integral explosion condition of \citet{murphy17} which is a generalization of
the foundational work of \citet{burrows93}. 

Our attempt is not the first to include turbulence in
  calculations of critical conditions.  \citet{yamasaki05} and
\citet{yamasaki06} considered several effects that might
  reduce the explosion condition, including rotation and convection.
  However, their attempt to model turbulence only included a term representing
  turbulent enthalpy flux.  A more self-consistent derivation of a
  turbulence model should include three terms at a minimum: the
  turbulent enthalpy flux, the turbulent ram pressure, and turbulent
  dissipation. Without including all of these effects, it is unclear how turbulence
would actually effect the critical condition for explosion.  In this
manuscript, we incorporate the turbulence model of \citet{murphy13}
which includes these three terms and has been validated with
  core-collapse simulations. 

First, in subsection~\ref{critcurve}, we revisit criticality and present the relevant equations and
  assumptions. Next, in section~\ref{RD}, we present our methods for incorporating
  turbulence into these equations. To
  find the solutions for both the average background flow and the
  turbulent flow, we decompose the equations into the average and turbulent quantities, commonly called Reynolds decomposition.  By decomposing the
    variables into average and turbulent flows, we introduce extra
    variables.  To close the system of equations, we motivate a
    turbulence model in~\ref{Turbulence}, and we specify our
    assumptions. In~\ref{turbmethod} we describe our method for
    solving this system of equations. As it turns out, there
      is maximum allowable Reynolds stress in the stalled shock
      solutions.  Even though this upper limit does not affect the
      critical condition, we derive an analytic expression for it in
      section~\ref{Rcap} and include it in our critical condition
      calculations. Finally, in section~\ref{Parameters}, we discuss
    some of our assumptions and parameters
    that need to be verified by multi-dimensional
      simulations

\subsection{The Critical Curve}
\label{critcurve}

  Two explosion conditions which start from first
  principles are the critical curve of \citet{burrows93} and
  a generalization of this condition, the
  $\Psi$ condition of \citet{murphy17}. We will modify these critical
  conditions to include turbulence, so first we revisit what constitutes
  a critical condition. 

\citet{burrows93} introduced a critical curve
  that divides steady-state solutions from no solutions in the
  $L_\nu-\dot{\mathcal{M}}$ plane.  Below this curve, one can
  find steady-state stalled shock solutions that satisfy all boundary
  conditions.  Above the curve, there are no stalled-shock solutions which satisfy
  all of the boundary conditions. In detail, the main
  discriminant for finding steady-state solutions is whether or not
  they could match the density at the neutrinosphere with the stalled
  shock solution. They suggested, but did  not prove, that the
  solutions above the $L_\nu-\dot{\mathcal{M}}$ curve are explosive. 
  
  \citet{murphy17} took one step closer in proving that the solutions above the
  critical curve are explosive. They showed that the only steady
  solutions above the curve have $v_s > 0$, lending support to the supposition of
  \citet{burrows93}. They did this by connecting the discriminant of
  \citet{burrows93} (the $\tau = 2/3$ condition) to a new dimensionless
  parameter $\Psi$. This $\Psi$ parameter is a measure of overpressure
  compared to hydrostatic equilibrium and directly indicates
    whether the shock would move out, in, or
  stay stationary. Because it is more straightforward, we do our calculations in the \citet{burrows93}
formalism, using the neutrino density discriminant, but we will also present our results in the context of
$\Psi$, the overpressure. While the method of \citet{burrows93} is
simpler, \citet{murphy17} provides a more direct connection to $v_s$, and thus explosion.
  
The first step in finding the critical curve is finding
steady-state solutions to the Euler equations. The steady state equations are
\begin{equation}
\nabla \cdot (\rho u) = 0 \, ,
\label{mass}
\end{equation}
\begin{equation}
\nabla \cdot (\rho u \otimes u) = -\nabla P - \rho\nabla\Phi  \, ,
\end{equation}
and
\begin{equation}
\nabla\cdot \left [\rho u \left (h + \frac{u^2}{2} \right ) \right ] = - \rho u \cdot \nabla\Phi +\rho q \, .
\label{energy}
\end{equation}
Where $\rho$ is mass density, $u$ is velocity, $P$ is
  pressure, $\Phi$ is gravitational potential, $h$ is enthalpy, and $q$ is
  the total heating. In general, heating and cooling by
neutrinos is best described by neutrino transport \citep{janka17a,tamborra17}; we simplify neutrino transport by invoking a
simple light-bulb prescription for neutrino heating and a local cooling \citep{janka01}
\begin{equation}      
q = \frac{L_\nu \kappa}{4\pi r^2} - C_0 \left (\frac{T}{2 MeV}\right )^6 \, .
\end{equation}
I.e., we've adopted a spherically symmetric neutrino source. r is the radius from this source, $L_\nu$ is the neutrino luminosity emitted from the core of the star, and
  \begin{equation}
  \kappa = 7.5 \times 10^{-28} \left ( \frac{T_\nu}{4 MeV} \right )^2 (Y_n + Y_p) [g \cdot cm^2]
  \end{equation}
   is the neutrino opacity \citep{murphy13}; where $Y_n$ and $Y_p$ are the neutrino and proton fractions per baryon. $\kappa$ is related to the optical depth by
\begin{equation}
\tau = \int\limits_{r_g}^{r_s} \rho \kappa dr \, ,
\end{equation}
which dictates the absorptivity of the material in the gain region; $r_s$ and $r_g$ are the positions of the shock and gain radius, respectively. T is the
  matter temperature, and $C_0$ is the cooling factor ($1.399 \times
  10^{20}$ ergs/g/s). 
  
Following in the steps of \citet{burrows93} equations~(\ref{mass}-\ref{energy}) represent a boundary value problem with the boundaries being the neutron star surface and the shock. At the shock, we want the pre-shock, inflowing material to match the post-shock material through the Rankine-Hugoniot jump conditions. For the moment, let us assume that $v_s$ is zero, and in this case, the jump conditions become:
\begin{equation}
\rho_1 u_1 = \rho_2 u_2 \, ,
\label{masss}
\end{equation}
\begin{equation}
\rho_1 u_1^2 + P_1 = \rho_2 u_2^2 + P_2 \, ,
\end{equation}
and
\begin{equation}
\rho_1 u_1 \left (e_1 + \frac{1}{2} u_1^2 + \frac{P_1}{\rho_1} \right ) = \rho_2 u_2 \left (e_2 + \frac{1}{2} u_2^2 + \frac{P_2}{\rho_2} \right ) \, .
\label{energyy}
\end{equation}
Where $e$ is the internal eneregy and the subscripts 1 and 2 indicate the downstream and upstream flows at the jump, respectively.
Normally, one uses eq.~(\ref{masss}) in eq.~(\ref{energyy}) to eliminate the
  mass flux in the Hugoniot-Rankine jump condition.  Here, we explicitly
  include it because this term can not be neglected when we Reynolds
  decompose and derive the
  jump conditions including the turbulent terms.
One then integrates inward to the neutron star surface where the
density profile must match the neutron star surface such that the
neutrino optical depth is $2/3$. If the neutrino optical
  depth is not $2/3$, then one searches for a new $r_s$ so that
the shock and neutron star boundary conditions are met.  In
  practice, \citet{yamasaki05} noticed that the density	at the neutrinosphere has about the same value in most
	situations.  For the opacities used in this manuscript,
  \citet{murphy08b} calculated the density in one-dimensional
  and two-dimensional simulations to be about $7 \times 10^{10}$ g cm$^{-3}$ at
the neutron star ``surface.''  Since this density condition is faster
to integrate, we use it instead of the neutrino optical depth condition.
  
While \citet{burrows93} provide an elegant one-dimensional explosion
condition, it does not accurately diagnose realistic supernova
explosions in multiple dimension \citep{murphy08b}. However, subsequent multi-dimensional work
suggests that one might be able to augment the technique for finding
the critical curve to include turbulence \citep{murphy13,murphy11,hanke12,couch13,radice16}. To do this, we build on the work of
\citet{murphy11} and \citet{murphy13}, and use the Reynolds decomposed continuity equations
to find a multi-dimensional critical curve. 

\subsection{Reynolds Decomposed Equations}
\label{RD}


A standard method to incorporate turbulence is through Reynolds
decomposition. The primary goal is to derive mean-field,
  steady-state equations for turbulence.  The
  first step is to Reynolds decompose the variables into background
  and perturbed components; i.e. $u = u_0 + u^{\prime}$,
    where $0$ denotes the background component and the prime indicates
    the turbulent term.  Next, one substitutes these terms
	  into the time-dependent Navier-Stokes equations.  To
	  obtain the mean-field correlations for turbulence, one averages the equations both
  in time and solid angle. For simplicity,
	we denote both of these averages by the operator $\langle \cdot
  \rangle$.  Since $\langle u \rangle = u_0$ and $\langle
  u^{\prime} \rangle = 0$, terms that involve one component of a
  turbulent variable are zero.  All non-zero turbulent terms are higher
order correlations of turbulent variables.  

Technically, the turbulence represents a
	time-dependent fluctuation.  However, the mean-field variables, or
  turbulent correlations are time-averaged correlations and can be in
  steady-state.  For core-collapse simulations, the turbulent
  correlations are effectively in steady-state, so one may drop the
  time derivatives in the Reynolds-averaged equations.  The resulting
  equations represent steady-state equations for the background flow
  and the mean-field turbulent correlations.  In this manuscript, we
  highlight the important correlations and steady-state equations.
  For a more thorough derivation of the equations from the
  Navier-Stokes equations, see \citet{meakin07b} or \citet{murphy11b}.

The three
dominant Reynolds turbulent correlations are the Reynolds Stress
($\textbf{R}$), turbulent dissipation ($\epsilon_k$), and turbulent
luminosity ($L_e$) \citep{murphy11,murphy13}. The Reynolds stress is
the turbulent fluctuation in momentum stress, turbulent luminosity is the transport of turbulent internal energy, and turbulent dissipation is the viscous conversion of mechanical
energy to heat. These terms are
\begin{equation}
  R_{ij} = u_i' u_j' \, ,
\end{equation}
\begin{equation}
  L_{e,i} = 4 \pi r^2 \rho_0 \langle u_i' e' \rangle = 4 \pi r^2 F_{e,i}  \, , 
\end{equation}

and in the limit of small viscosity, turbulent dissipation is
\begin{equation}
  {\bm \epsilon} = 2 \nu ( \nabla u') \cdot ( \nabla u' ) \, .
\end{equation}

The turublent kinetic energy dissipation is $\epsilon_k = tr({\bm \epsilon})/2$. Note that this definition is slightly different from the \citet{murphy11b} definition which presented a confusing sign and needlessly included turbulent diffusion in the turbulent dissipation term. Here, we take the limit of small viscosity so the turbulent diffusion term goes away, but turbulent dissipation remains. Furthermore, we corrected the sign so that a positive $\epsilon$ corresponds to taking energy from the kinetic energy equation and putting it in the internal energy equation.  Since this term requires higher order correlations, we model it using Kolmogorov's assumptions (see section \ref{Turbulence} or refer to the result of \citet{canuto93} for a more robust description).

The resulting steady-state, Reynolds-decomposed equations are
\begin{equation}
\nabla \cdot (\rho_0 \vec{u}_0 + \langle \rho' \vec{u}' \rangle) = 0 \, ,
\label{consmass}
\end{equation}
\begin{equation}
\langle \rho \vec{u} \rangle \cdot \nabla \vec{u}_0 = -\nabla P_0 + \rho_0 \vec{g} - \nabla \cdot \langle \rho \textbf{R} \rangle \, ,
\label{consmom}
\end{equation}
and
\begin{equation}
\begin{aligned}
  \langle \rho u \rangle \cdot \nabla e_0 + \langle P_0\nabla \cdot u_0 \rangle + \langle P' \nabla \cdot u' \rangle = \\ -\nabla \cdot F_e + \rho_0 q + \rho_0 \epsilon_k \, .
\label{consenergy}
\end{aligned}
\end{equation}
To see the exact equation, please refer to \citet{meakin07b}. There,
they have fully expanded the above equation into their background and
perturbed components. The internal energy flux and Reynolds stress are
\begin{subequations}
\begin{equation}
\label{eq:FeDefine}
F_e = \langle \rho u e' \rangle 
\end{equation}
and
\begin{equation}
\label{eq:Rdefine}
\langle \rho R \rangle = \langle \rho uu' \rangle \, .
\end{equation}
\end{subequations}
Alternative and common
  definitions are $F_e = \rho_0 \langle u' e' \rangle$ and $\langle \rho R \rangle = \rho_0 \langle  u'u' \rangle$.  We use eq.~(\ref{eq:FeDefine})
  because it gives the same result but is much
  simpler and cleaner to calculate in numerical simulations.
  Expanding equations~(\ref{eq:FeDefine}) and (\ref{eq:Rdefine}) gives
\begin{subequations}
\begin{equation}
F_e = \rho_0 \langle u' e' \rangle + \langle \rho' u' e' \rangle + u_0
  \langle \rho' e' \rangle 
\end{equation}
and
\begin{equation}  
\langle \rho R \rangle = \rho_0 \langle u' u' \rangle + \langle \rho' u' u' \rangle + u_0 \langle \rho' u' \rangle  \, .
\end{equation}
\end{subequations}
Within the convective region,
  \citet{murphy11b} and \citet{murphy13} found that the first term is
  the dominant term.  However, just using the first term creates a
  large spike at the aspherical shock which has nothing to do with
  convection and everything to do with the jump conditions across the aspherical
  shock.  However, using eq.~(\ref{eq:FeDefine}) mitigates this
  problem and gives the correct turbulent energy flux within the
  convective region.  

The Decomposed boundary conditions are:
\begin{equation}
\rho_1 u_1^2 + P_1 + \rho_1 R_{rr} = \rho_2 u_2^2 + P_2 
\label{demomentum}
\end{equation}
and
\begin{equation}
  \frac{P_1}{\rho_1} + e_1 + \frac{L_e}{\mdot} + \frac{L_k}{\mdot} + \frac{1}{2}u_1^2 + R_{rr} = \frac{P_2}{\rho_2} + e_2 + \frac{1}{2}u_2^2 \, ,
\label{deenergy}
\end{equation}
where $\dot{\mathcal{M}} = 4 \pi r^2 \rho u$ is the mass accretion rate, $L_e = 4 \pi r^2 \rho_0 \langle u' e' \rangle$ is the internal energy luminosity, and $L_k (= \rho_0 u_0 \langle u'^2/2 \rangle)$ is the kinetic energy luminosity. 
  Now that we have introduced three new
turbulent variables, we have a total of six unknown variables and only
three equations, necessitating more equations.

\subsection{Turbulence Models}
\label{Turbulence}

%


Including the turbulent components, the Reynolds-decomposed conservation equations,
(\ref{consmass}-\ref{consenergy}), now have more variables than
equations. These extra variables are the
Reynolds stress, turbulent luminosity, and turbulent
dissipation.  Therefore, to find a solution to these equations, we need a
  turbulence closure model.  Turbulence depends upon the bulk
  macroscopic flow, so the equations for turbulence represent a
  boundary value problem that depends upon the specifics of the
  background flow.  For this reason, \citet{murphy13} developed a
  global turbulence model for neutrino-driven convection.  The
  steady-state equations require local turbulent expressions and
  derivatives.  Therefore, to use the global model, we must
  make some assumptions and translate the global model to a local model.

In the core-collapse problem, there may be two
sources of turbulence: convection and the SASI \citep{bethe90,blondin03}. In
  principle, to correctly model turbulence in the core-collapse
  problem, we need a turbulence model that addresses both driving
  mechanisms: convection and the SASI.  While there are nonlinear
  models to describe turbulent convection, there are no nonlinear models yet to
  describe SASI turbulence.  Thus, we proceed with a
convection based analysis of turbulence. 

There are five turbulent variables
  ($\textbf{R},L_e,\epsilon_k$), three of them are Reynolds stress
  terms ($R_{rr}$, $R_{\phi \phi}$, and $R_{\theta \theta}$);
  therefore, we invoke five constraints.
Our five global constraints are as follows. First,
    we eliminate the tangential components of the Reynolds stress.  In
  neutrino-driven convection, the radial Reynolds stress is in rough equipartition with both of the tangential components \citep{murphy13}:
\begin{equation}
R_{rr} \sim R_{\phi \phi} + R_{\theta \theta} \, .
\label{firstr}
\end{equation}
Similar simulations showed that the transverse components are roughly the same scale:
\begin{equation}
  R_{\phi \phi} \sim R_{\theta \theta}
  \label{secondr}
\end{equation}
Using \citet{kolm}'s hypothesis we relate the Reynolds stress
to the turbulent dissipation: 
\begin{equation}
\epsilon_k \approx \frac{u'^3}{\mathcal{L}} \ =
\frac{R_{rr}^{3/2}}{\mathcal{L}} \, ,
\label{second}
\end{equation}
where $\mathcal{L}$ is the largest turbulent dissipation scale.
From \citet{murphy11}, we note that buoyant driving roughly
  balances turbulent dissipation:
\begin{equation}
W_b \approx E_k \, ,
\label{third}
\end{equation}
where the buoyant driving is the total work done by buoyant
  forces in the convective region,
\begin{equation}
W_b  = \int\limits_{r_g}^{r_s} \langle \rho' u'_i \rangle g^i dV \, ,
\label{wb}
\end{equation}
and the total power of dissipated turbulent energy is
\begin{equation}
\label{eq:Ekdefine}
E_k = \int\limits_{r_g}^{r_s} \rho \epsilon_k dV \, .
\end{equation}
Lastly, three-dimensional simulations from \citet{murphy13} show that the source of neutrino-driven convection, the neutrino power, is related to the turbulent dissipation and the turbulent luminosity by
\begin{equation} 
L_\nu \tau \approx E_k + L_e^{max} 
\label{first}
\end{equation}
Together, equations~(\ref{firstr}-\ref{third}) and (\ref{first}) represent our turbulence closure model.

Now that we have combined a series of global conditions to close our global model, we must relate these back to local functions in order to incorporate them into our conservation equations. We do this by making assumptions about the local profile for each term and scaling them by one parameter for each turbulent term. In translating from global to local, we introduce three parameters of the turbulent region: a constant Reynolds stress ($\textbf{R}$), a constant dissipation rate ($\epsilon_k$), and a maximum for the turbulent luminosity ($L_e^{max}$); the corresponding local terms are $\nabla \cdot \langle \rho \textbf{R} \rangle$, $\rho_0 \epsilon_k$, and $\nabla \cdot \langle \vec{L}_e \rangle$ respectively (see equations (\ref{consmom}-\ref{consenergy})). Thus, the final solution for turbulence boils down to finding these three parameters.

To find the three parameters, we insert the profiles for the turbulent
terms and their parameters into the global conditions,
equations~(\ref{firstr}-\ref{third}) and (\ref{first}). This then leads to
a set of equations for the parameters, and we use simple
algebra to solve for the scale of the parameters that satisfies those
global conditions. We solve for these in the order they are presented: the Reynolds stress, turbulent dissipation, and finally the turbulent luminosity. 

Our first task is to reduce the three Reynolds stress terms
  down to one. In neutrino-driven convection, there is a preferred direction (i.e. in
the direction of gravity) and simulations show that there is an
equipartition between the radial direction and both of the tangential
directions \citep{murphy13}. Simulations also show that the two
tangential directions have the same scale \citep{murphy13}. Evaluating
these assumptions in equations~(\ref{firstr}-\ref{secondr}) reduces the representation of the Reynolds stress as three variables down to one:
\begin{equation}
  R = 2R_{rr} \, .
\end{equation} 

Kolmogorov's theory of turbulence predicts the turbulent dissipation rate scales as the
perturbed velocity cubed over the characteristic length of the
instabilities, equation~(\ref{second}). Numerical simulations suggest
  that the largest dissipation scale in convection is the size of
  the convective zone \citep{couch14,fogl15,fern15}, or the gain region in the core-collapse case. Moreover, eqs.~(\ref{third}-\ref{first})
have a global definition of $\epsilon_k$, and so we assume that
$\epsilon_k$ is roughly constant over the gain region. Therefore, from equation~(\ref{eq:Ekdefine}),
\begin{equation}
\label{Ek}
  E_k = \int_{r_g}^{r_s} \rho \epsilon_k dV \, ,
\end{equation}
we have
\begin{equation}
  \epsilon_k \approx \frac{E_k}{M_{gain}} = \frac{W_b}{M_{gain}} \, .
  \label{epk}
\end{equation}

The final missing piece is the Turbulent Luminosity, $L_e$, and we
connect this to the turbulent dissipation by rewriting the buoyant
work in terms of the turbulent luminosity. We combine the observation that
buoyant work is approximately equal to the turbulent dissipation power
\citep{murphy13} and that this energy can be converted to heat
\citep{murphy11}.  
Ignoring compositional perturbations, the density perturbation in terms of
the perturbed energy and pressure is 
\begin{equation}
  \rho ' = e'\big(\frac{\partial \rho}{\partial e}\big)_P + P'\big(\frac{\partial \rho}{\partial P}\big)_e \, .
\end{equation}
Convective flows are generally dominated by buoyant
  perturbations and not pressure perturbations.  Even for high mach
  number convection, buoyancy tends to dominate; see
\citet{murphy13}.  Therefore, one may
express the density perturbation in terms of the energy perturbation alone. Applying the above step and some algebra to (\ref{wb}), we obtain
\begin{equation}
\begin{aligned}
W_b = \int \frac{ (\gamma - 1)L_e \Phi \rho }{P} \frac{dr}{r} \, .
\end{aligned}
\end{equation}
Now that we have an equation for the buoyant driving as a function of
the turbulent luminosity ($L_e$), we now need the radial profile for $L_e$
  to complete the connection between turbulent luminosity and buoyant
  driving.  Initial numerical investigations of $L_e$ suggest that it
  rises from zero at the gain radius to a nearly constant value above
  the gain radius until the shock \citep{murphy13}.  Therefore, we develop an ansatz for
$L_e$ which roughly satisfies the shape seen in simulations
\begin{equation}
L_e = L_e^{max} \tanh\bigg(\frac{r-r_g}{\textbf{h}}\bigg) \, .
\label{le}
\end{equation}
Where $\textbf{h}$ is the distance it takes for the turbulent
  luminosity to increase from zero to roughly $L_e^{max}$. Thus, our buoyant driving power becomes
\begin{equation}
W_b  = (\gamma - 1)L_e^{max} \int \tanh \left ( \frac{r -
  r_g}{h} \right ) \frac{\Phi \rho}{ P } \frac{dr}{r} \, .
\end{equation}
 Finally, relating this back to our global condition (\ref{first}) and (\ref{epk}), we can find the maximum turbulent luminosity
\begin{equation}
L_e^{max} \approx \frac{L_\nu \tau}{1 + (\gamma - 1) \int
  \tanh \left ( \frac{r - r_g}{h} \right ) \frac{\Phi \rho}{ P } \frac{dr}{r}}
\, .
\label{lemax}  
\end{equation}

Within the context of our assumptions, equation~(\ref{lemax}) is our
final algebraic expression which gives the scale of turbulence in
terms of the background structure and the driving neutrino power.  The
second term in the denominator is a weak function of shock radius and
is of order unity.  Therefore, the turbulent luminosity is a fraction
of the neutrino-driving power $L_{e}^{max} \approx L_{\nu} \tau / (1 +
f(R_s))$, and the turbulent dissipation is also a fraction of the
driving neutrino power $E_k \approx L_{\nu} \tau f(R_s)/ (1 +
f(R_s))$.

Though we have successfully closed our set of equations, we did employ
several assumptions about the local structure of turbulence. Many of the assumptions made in this section have been
verified by simulations individually \citep{murphy13} and, since we
have been careful in being self-consistent, should have equal validity
when combined. However, there are some assumptions that require
further verification, and in section~\ref{turbmethod}, we discuss
these details.

\subsection{Solution Method Including Turbulence}
\label{turbmethod}

We seek steady-state solutions for the background flow including turbulence equations~(\ref{consmass}-\ref{consenergy}).  
   This is a global boundary value problem, and
  in this section, we describe our solution strategy.  In essence, the
equations for the turbulence model represent a boundary value problem
for turbulence
embedded within the larger boundary-value problem of the background
solution.  As is standard, to find the steady-state solutions
including turbulence, we represent this boundary value problem as a set
of coupled first order differential equations and use the shooting method
to find the global solution.

To "shoot" for our solution, we first designate the boundary
conditions at the shock for temperature, pressure, and density.  We then integrate inward to the neutron star surface and apply our final boundary condition.  To find the
  steady-state stalled solution, the density at the inner boundary
  should satisfy $\tau = 2/3$.  In the cases that the density does
  not match the inner boundary, then these ``solutions'' represent
  pseudo steady-state solutions, and the ratio of the inner density to
  the desired inner density provides an indication of whether the
  shock would move out or in (see equations (18-21) of \citet{murphy17}).
  
Now that we have suitable boundary conditions, we can find the cases
for which steady-state solutions exist. Since we have several free
physical parameters ($L_\nu$, $\dot{\mathcal{M}}$, $M_{NS}$, $R_{NS}$, $T_\nu$), we fix three of them at reasonable values, iterate over a fourth parameter, and calculate
the critical value of the fifth which satisfies the tau condition. For example, in the spirit of \citet{burrows93}, we fix $M_{NS}$, $R_{NS}$, $T_\nu$, vary $\dot{\mathcal{M}}$, and find a critical $L_\nu$. As is detailed in \citet{murphy17}, this critical curve
  corresponds to where $\Psi_{\rm min} = 0$ (which also corresponds to
  $v_s = 0$). 
  
  Previously, calculating this critical point only required local terms in the equations. Since our turbulence model relies on global quantities ($M_g$, $W_b$, $\tau$), we must specify realistic values for these global quantities initially, then use the density and temperature profiles from the previous iteration to calculate them for the following pseudo-solutions. This method is a valid approximation since the boundaries and constituents of each integral varies a negligible amount after each step.

\subsection{A New Upper Bound on the Reynolds Stress}
\label{Rcap}
We have discovered that there is an upper bound on the Reynolds stress
in the core-collapse problem. This comes from the fact that the
Reynolds stress applies a turbulent ram pressure at the shock, and for
large enough shock radii, there is an upper limit to the Reynolds
stress that allows solutions to the stalled shock jump conditions. Of course, the Reynolds stress depends upon the driving neutrino power,
but otherwise, we find that as one increases the shock
radius, the Reynolds stress only goes up slightly; however, the scale
of the ram pressure at the shock is set by the
gravitational potential energy, which decreases as 1/r. Eventually,
the ratio of the Reynolds stress to the gravitational potential
becomes so large that there are no longer solutions to the steady-state jump conditions. We now derive the analytic upper bound to the Reynolds stress and describe our implementation of this limit into our solution method.

To quantify this upper bound, we start with our three jump conditions
(\ref{masss}-\ref{energyy}) that include
turbulence, and make some assumptions so that we can derive
  an analytic expression. Our first assumption is that the fluxes of internal energy and kinetic energy at the shock are small relative to the other terms \citep{meakin07b,meakin10,murphy11b}. Our new decomposed boundary conditions are thus:
\begin{equation}
  P_1+\rho_1u_1^2+\rho_1R_{rr} = P_2 + \rho_2u_2^2
\end{equation}
and
\begin{equation}
  \frac{P_1}{\rho_1} + e_1 + \frac{1}{2}u_1^2 + R_{rr} \approx \frac{P_2}{\rho_2} + e_2 + \frac{1}{2}u_2^2
\end{equation}
Here we have omitted the 0 subscript, where all non-perturbed terms are implied to be the background. Since all of the perturbed components have either canceled or been defined, there is no need to differentiate with a $_0$ or $^\prime$

Using a $\gamma$-law equation of state we can solve for a solution of the ratio of densities:
\begin{equation}
 \beta = \frac{\gamma + 1}{\gamma + \mathcal{M}_2^{-2} - \sqrt{(1-\mathcal{M}_2^{-2})^2 - (\gamma+1)\frac{2R_{rr}}{u_2^2}}}
\end{equation}
Where $\beta$ is the compression factor, $\beta = \rho_1/\rho_2$. For physical solutions of $\beta$, the term under the radical needs to be positive. This sets an upper limit on the second term which is an upper limit on the Reynolds stress.  

We now make some approximations to derive a simple, analytic limit for the Reynolds stress. If we assume that the velocity of in-falling matter onto the shock is roughly in free fall:
\begin{equation}
  \frac{1}{2}v_2^2 \approx \Phi(r_s)
\end{equation}
an upper limit on our Reynolds stress becomes
\begin{equation}
  R_{rr} \leq \frac{(1-\mathcal{M}_2^{-2})^2}{\gamma + 1}\Phi(r_s) \, ,
  \label{Rrr}
\end{equation}
or in dimensionless form:
\begin{equation}
  \mathcal{R} = \frac{(\gamma+1)R_{rr}}{\Phi(r_s) (1-\mathcal{M}^{-2})^2} \leq 1 \, .
  \label{Rcapeq}
\end{equation}
Above this limit, there can be no stalled shock solutions, and since our method takes this assumption as an intermediate step, above this limit, we can not find these quasi-steady solutions. Thus, we terminate our $\Psi$ curve at the radius for which our $\mathcal{R}$ parameter crosses this threshold. In practice, we had numerical difficulty when $\mathcal{R}$ got close to one. So to avoid that numerical difficulty, we set a cap of 0.6 of this value (see fig. \ref{manycaps}).

This upper limit on the Reynolds stress could have affected the
critical curve, however it does not. To reiterate: the critical
neutrino luminosity curve is determined by the point of the minimum of
the $\Psi$ curve. Theoretically, if the imposed upper limit on
$\mathcal{R}$ were to be at a shock radius smaller than our $\Psi_{\rm
  min}$, an entirely new critical curve might have to be defined in
order to ensure that the above constraint was not
being violated. Luckily, in all cases where an actual cap is
necessary, this happens to be at a shock radius greater than
$\Psi_{\rm min}$. Thus, the cutoff point that we use, $\sim$60\%, is
synonymous with the condition of Eq.~(\ref{Rrr}). Though imposing an
upper limit on $R_{rr}$ would ostensibly mitigate its affect on the
critical curve, we have just shown that all pseudo-solutions after
$\Psi_{\rm min}$ are irrelevant. Hence, we are only constraining the
Reynolds stress for pseudo-solutions which are already non-steady
state. That said, it is possible that once we start looking at the
\textit{explodability} of a set of initial conditions, our upper limit
may start to interfere with predictions. This upper limit will be treated accordingly with more rigor in future publications. 

\subsection{Discussion, Parameters, \& Limitations of the Method}
\label{Parameters}





The approach that we outline in this manuscript provides
  a unique way to investigate how turbulence affects the critical
  condition for explosion, but it will require further validation.  At the moment, this approach is more of a
  proof of concept; to make it more quantitative and predictive, there
  are some parameters and limitations that must be explored and
  calibrated with more realistic multi-dimensional simulations.  The
  primary parameters are associated with the size of the convective region, $\mathcal{L}$ in eq.~(\ref{second}), and the length scale for the turbulent
  luminosity, $\textbf{h}$ in eq.~(\ref{le}).

Since the relation between the Reynolds stress and turbulent dissipation is modified solely by the length scale of convection, it is imperative to treat $\mathcal{L}$ properly. \citet{kolm} argued  it to be the size of the largest eddies formed. Since neutrino-driven convection and SASI both exhibit eddies and sloshing on the order of the gain region \citep{couch14,fogl15,fern15,radice16}, taking the full length of the gain region is not disingenuous. Furthermore, simulations show that the inertial range scaling spans several orders of magnitude, so even the largest eddies should contribute appropriately to the conversion of kinetic energy to heat, assuming fast cascade times \citep{armstrong95}. 

Contrarily, there is little work done in developing an analytic turbulence model
for core-collapse, thus finding an appropriate length scale for which
the turbulent luminosity is relevant becomes another parameter of our
problem. For the sake of consistency, only one value of $\textbf{h}$
has been used throughout this paper. However, varying values of $\textbf{h}$ yield the same characteristic results (within realistic lengths). 

Moreover, since the majority of multi-dimensional effects are confined
to the gain region, we approximate the effects to be zero below the gain radius and above the shock. Additionally, simulations
have shown that the increase in entropy due to turbulence is seen to
be strictly within the gain region, further supporting our isolation
of the additional heating to the gain region \citep{murphy13}.

In general, we started with an integral model for
  turbulence, but for the solutions, we require local solutions and
  made some significant approximations.  For the most part, these
  approximations seem to be consistent with multi-dimensional
  simulations.  To validate these assumptions, the community will need
  to test these assumptions with multi-dimensional simulations.
  
\section{Results \& Discussion}
\label{section3}

\begin{figure}[t]
\epsscale{1.2}
\plotone{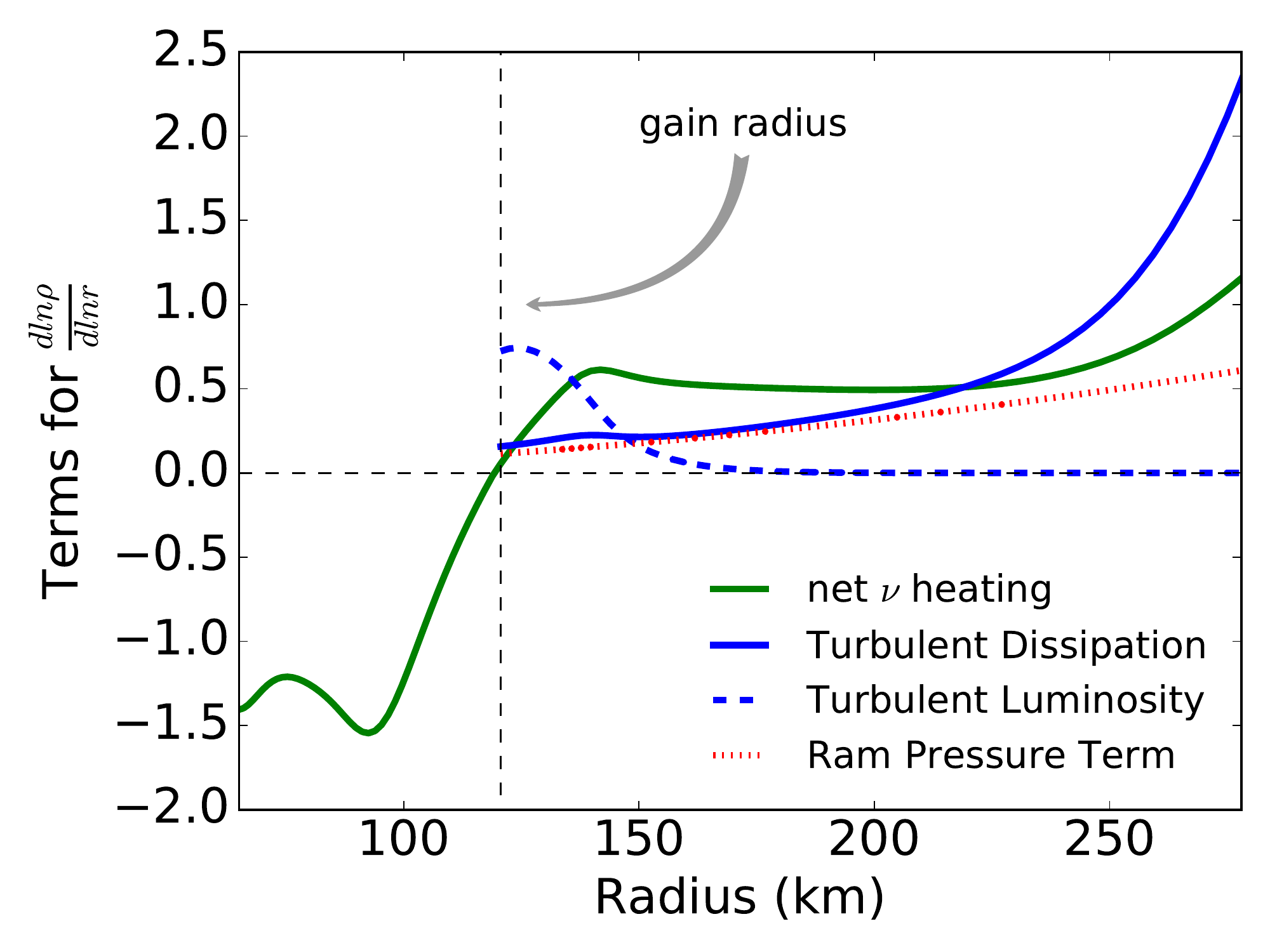}

\caption{The effects of neutrinos and turbulence on the density and temperature
  profiles.  Specifically, we show the dimensionless terms that appear
  in the equation for $\frac{d \ln \rho}{d \ln r}$, \ref{ODE1}. We omit the pressure and gravity terms
    which combine to give a power-law slope of about $-3$.  Instead we
  show the net neutrino heating and cooling term (solid-green line) and the effect of
  each turbulent term. The
  turbulent dissipation (solid, blue line) and the turbulent
  luminosity (dashed, blue line) terms originate
  from the energy equation, and the ram pressure term (dotted, red line)
  comes from the
  momentum equation.  In our model, we assume that turbulence is driven
  only in the gain region.  In general, the turbulent terms
  associated with the energy equation are larger than ram
  pressure.  More specifically, turbulent
  dissipation generally affects the profile more than the ram pressure. 
\label{normalized}
}
\epsscale{1.}
\end{figure}

\begin{figure}[t]
\plotone{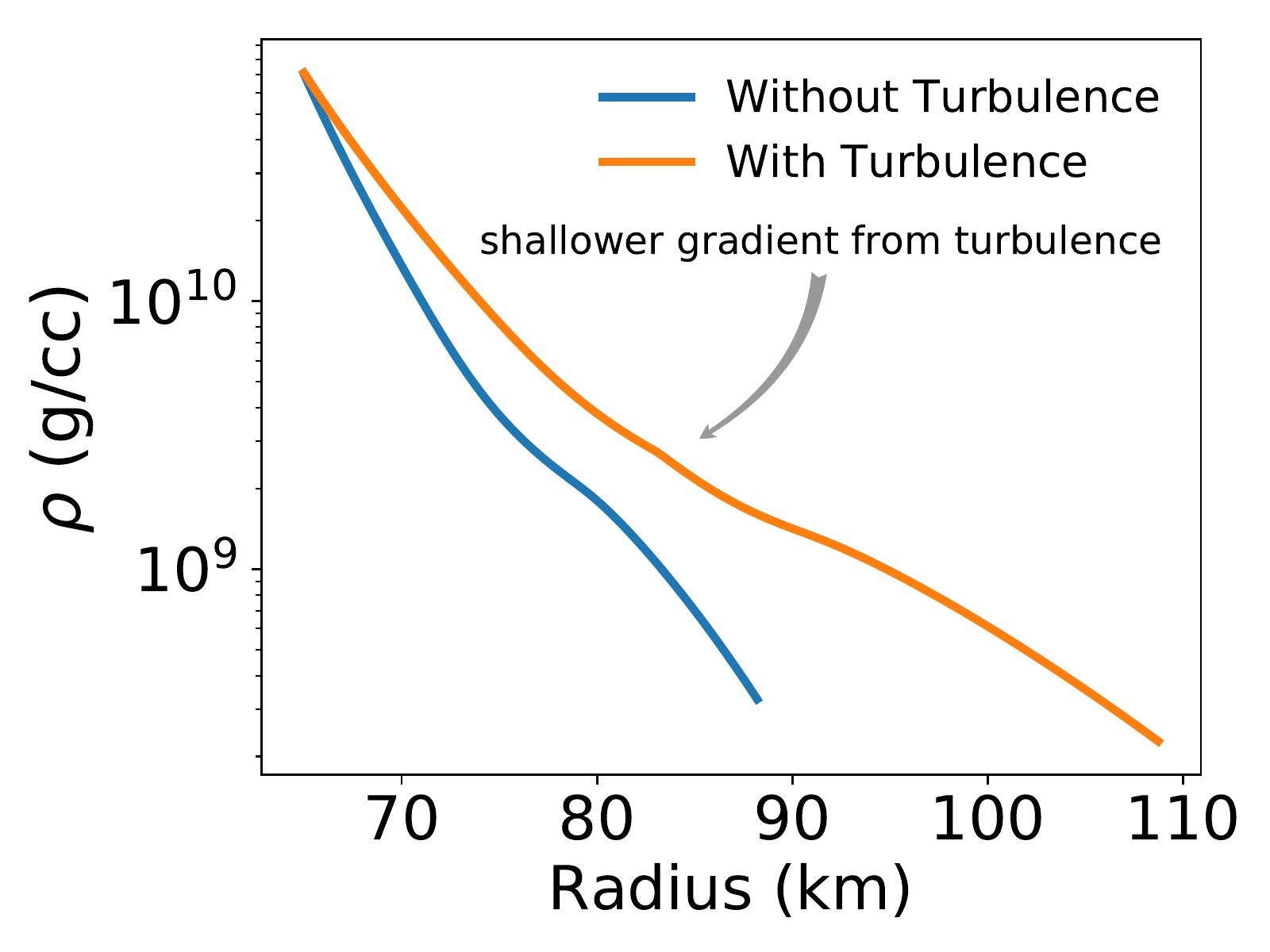}
\plotone{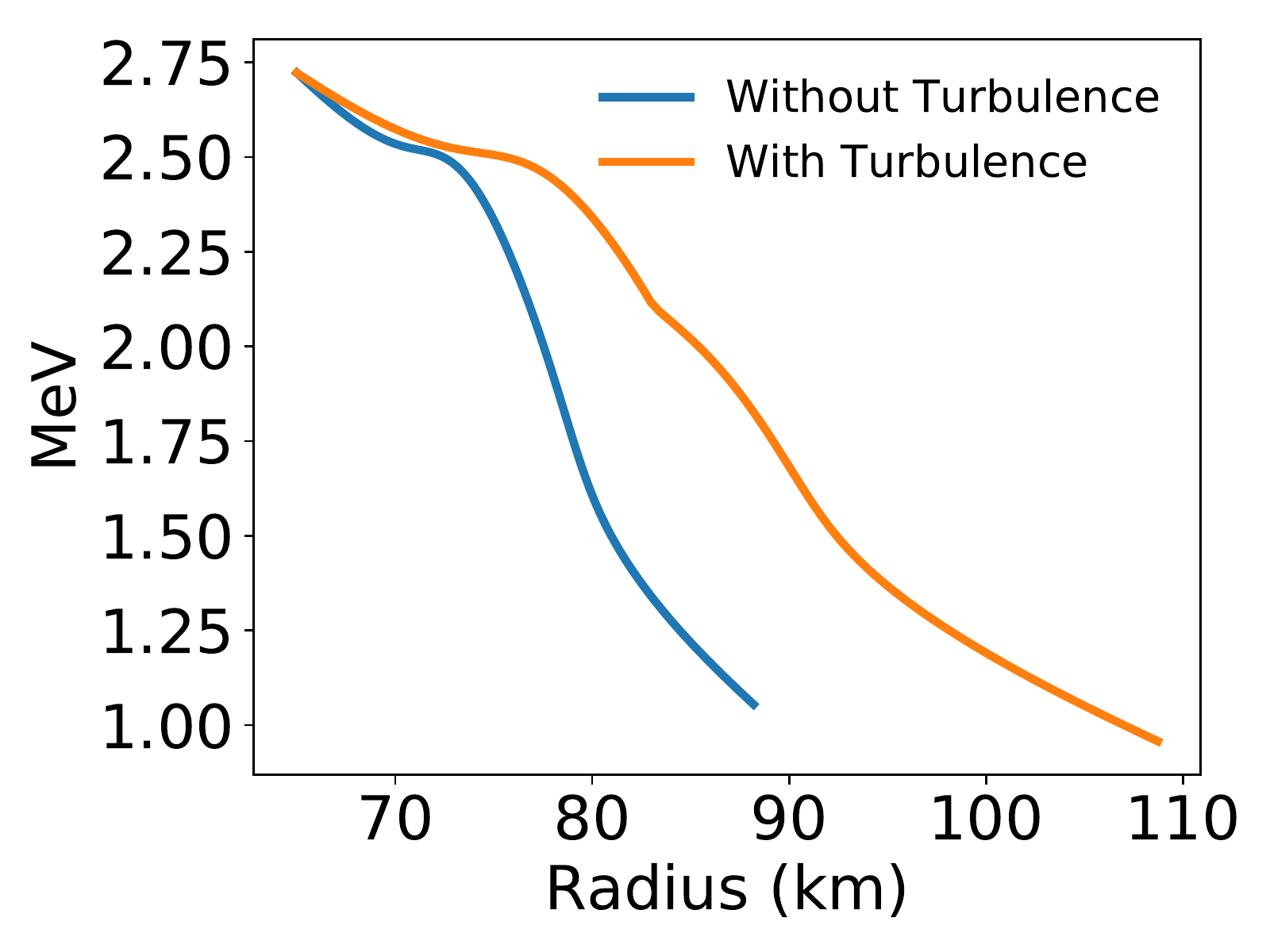}
\caption{The density and temperature profiles for stalled
    shock solutions with and without the convection. Generally, convection causes
    shallower gradients and higher temperatures.  While technically
    all of the terms contribute to this effect, the most dominant term
  is turbulent dissipation which provides extra heating.  
}
\label{denstemp}
\end{figure}

\begin{figure}[t]
\epsscale{1.2}
\plotone{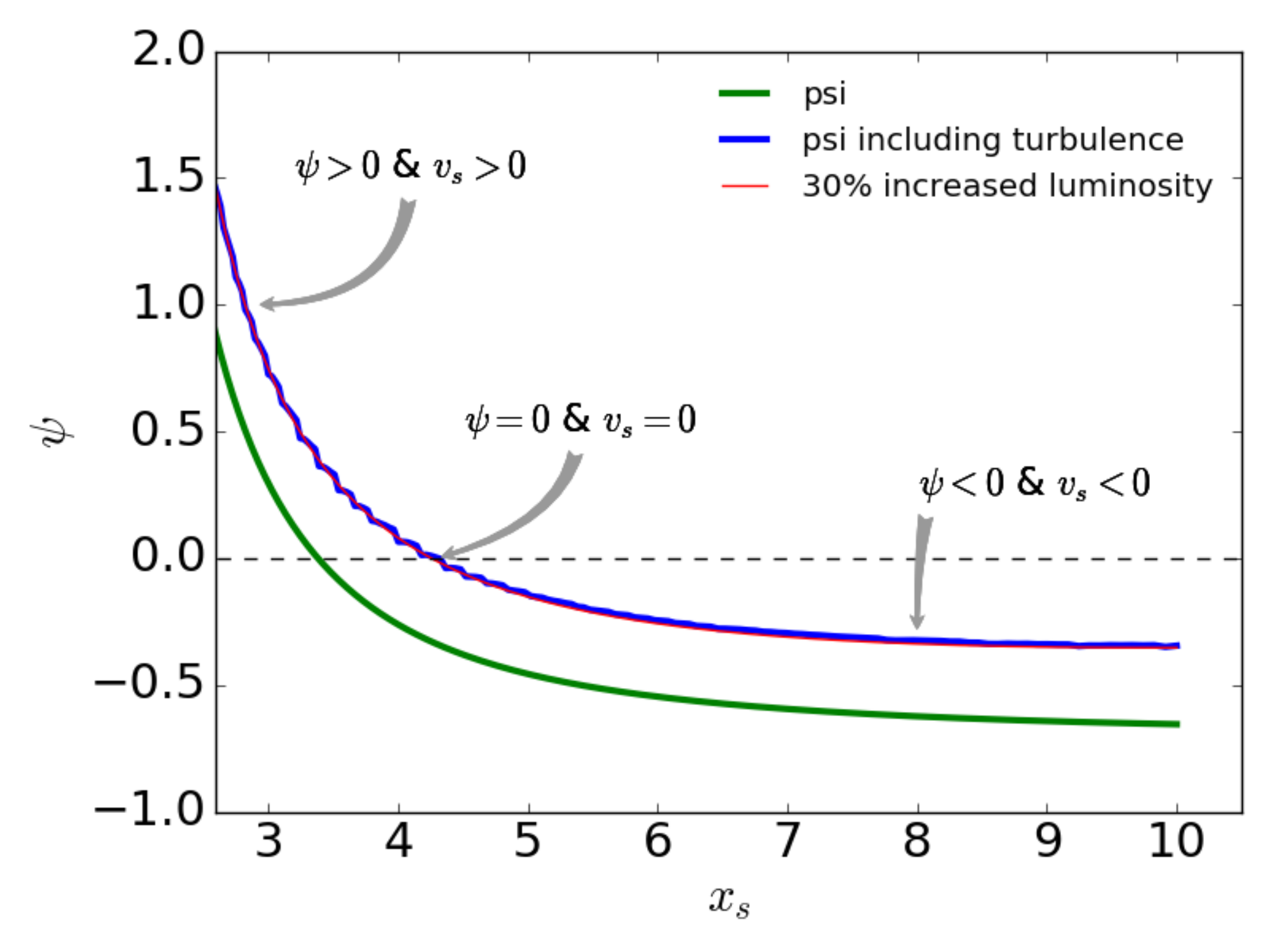}
\caption{$\Psi$ parameter as a function of
    shock radius, $x_s = R_s/R_{\rm NS}$.  $\Psi$ is roughly the over
    pressure compared to hydrostatic equilibrium normalized by the
    pre-shock ram pressure.  The shock velocity is related to
    $\Psi$ by $v_s \approx 1 - \sqrt{1 + \Psi}$.  Therefore, the sign
    of $\Psi$ determines whether the shock recedes, expands, or is
    stalled.  There is always a minimum for $\Psi$, $\Psi_{\rm min}$.
  If $\Psi_{\rm min} < 0$, then there is always a stalled solution.
  On the other hand, when $\Psi_{\rm min} > 0$, then only $v_s > 0$
  solutions exist.  \citet{murphy17} propose that $\Psi_{\rm min}=0$
  is the explosion condition.  Adding turbulence has the effect of
  raising $\Psi_{\rm min}$, making it easier to reach the explosion
  condition.  Adding turbulence has a similar effect as increasing the
  neutrino luminosity by 30\% (red line). 
}
\label{psi}
\epsscale{1.}
\end{figure}
\begin{figure}[t]
\epsscale{1.2}
\plotone{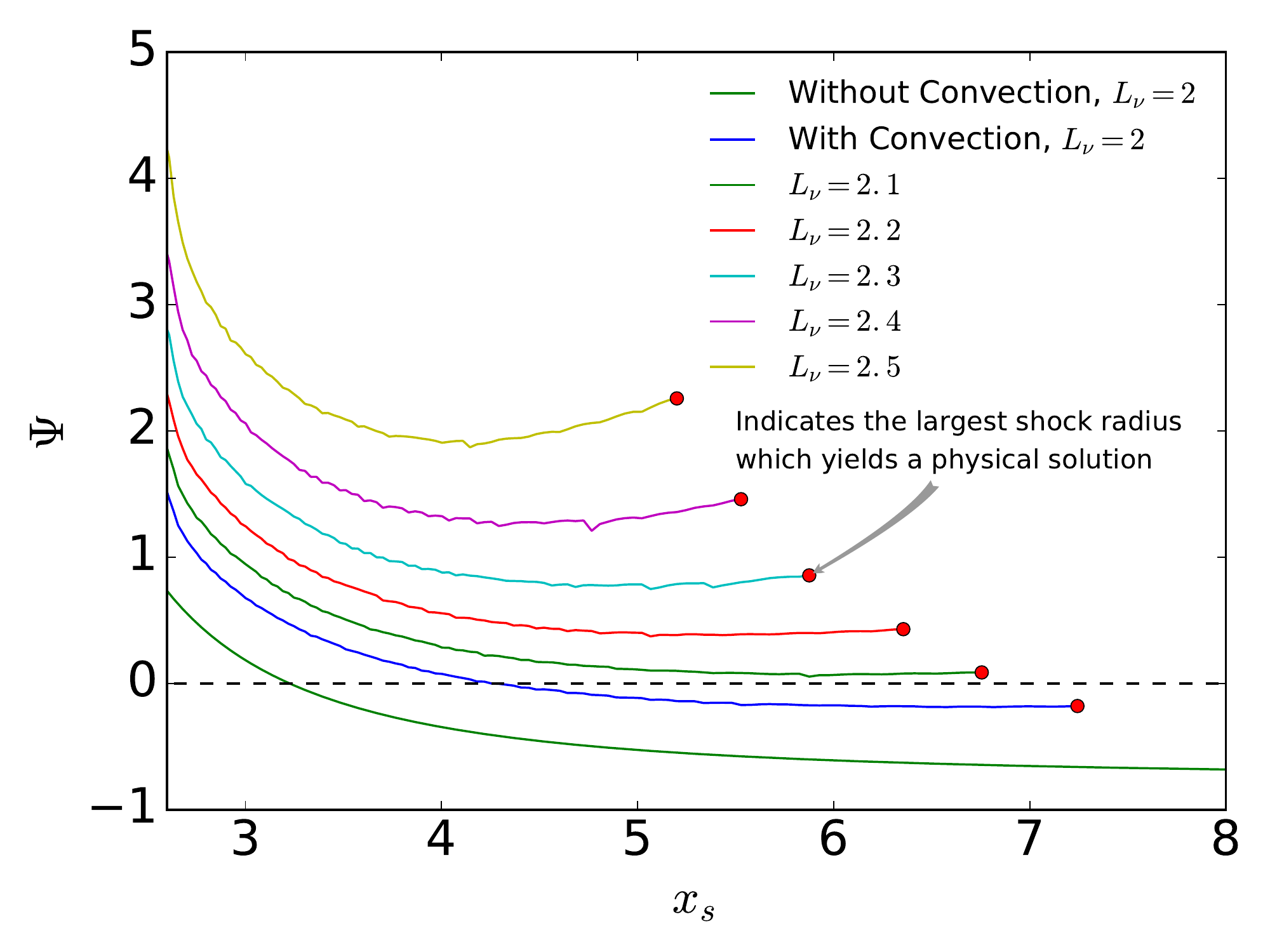}
\plotone{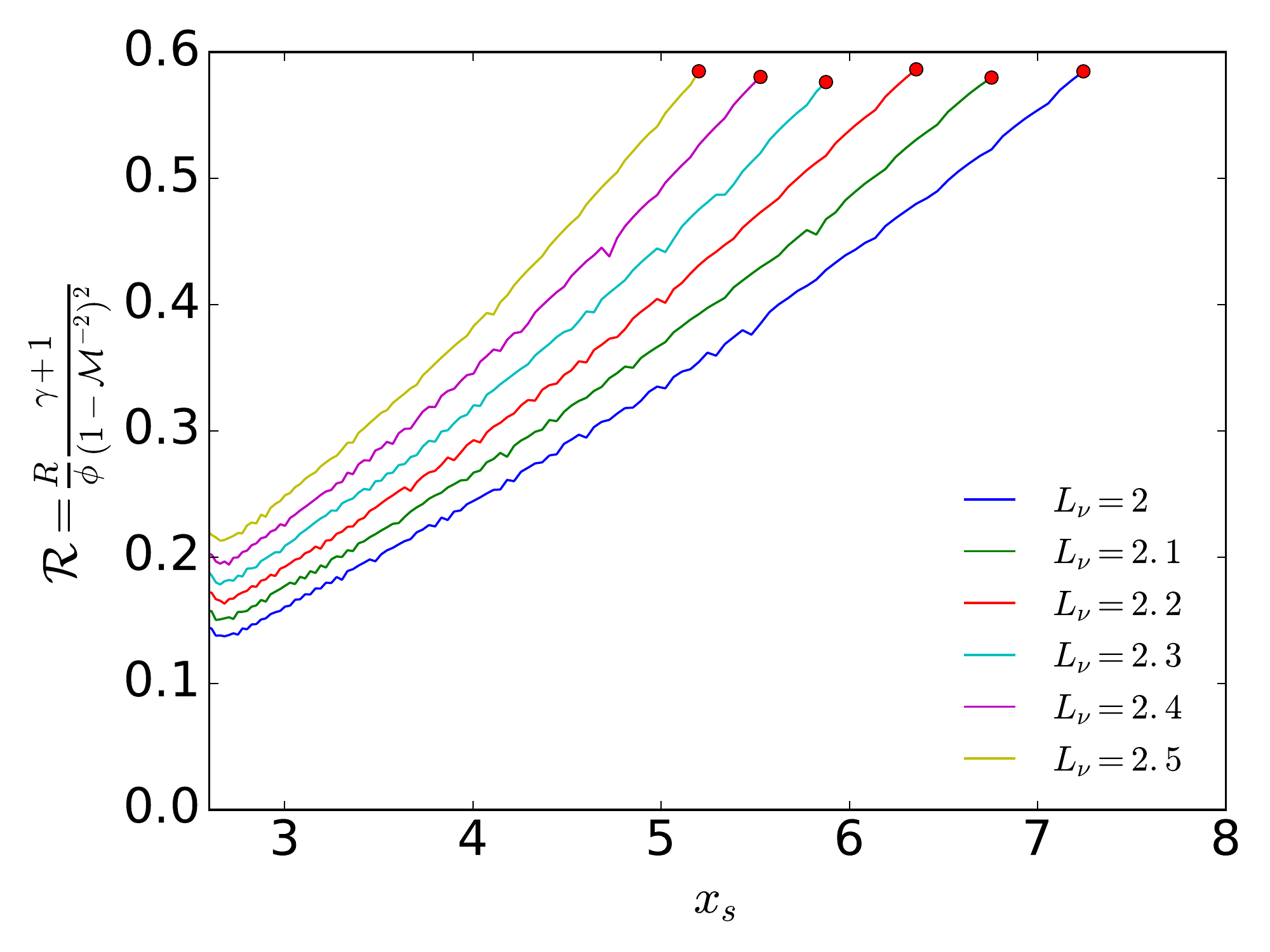}
\caption{$\Psi$ and corresponding dimensionless Reynolds stress, $\mathcal{R}$, as a function of shock radius ($x_s$ = r/$r_s$). There is an upper limit for the dimensionless Reynolds stress which we derive in section \ref{Rcap} (see equation (\ref{Rcapeq})). Here, we show the behavior of the normalized Reynolds stress vs. $x_s$ and how it affects the explodability parameter $\Psi$. In the bottom panel, each $\mathcal{R}$ increases until it reaches an unphysical point and terminates at the red dot. The same termination points can be seen above in the top panel, where each dot corresponds to its respective unphysical shock radius for a given $L_\nu$. If the cap had occurred to the left of $\Psi_{\rm min}$ then this upper bound on R would have affected the critical curve. However, the upper limit occurs to the right of $\Psi_{\rm min}$, therefore it does not affect the critical curve. Thus, the critical point of explosion is still dominated by non-ram pressure terms.}
\label{manycaps}
\end{figure}
\epsscale{1.}
\begin{figure*}[t]
\epsscale{1.3}
\plotone{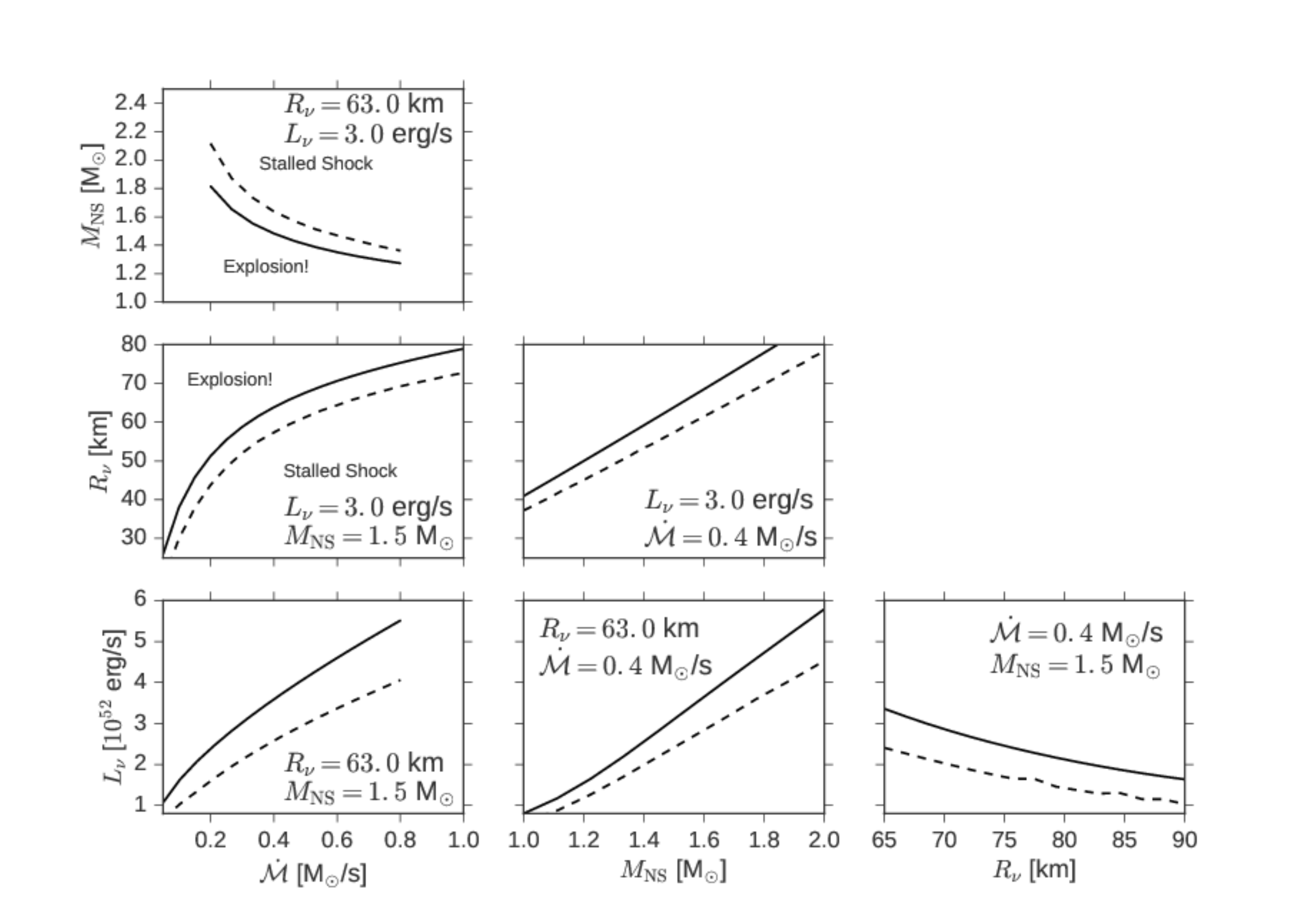}
\caption{How turbulence affects the $\Psi_{\rm min} = 0$
    explosion condition.  $\Psi_{\rm min}$ depends upon five
    parameters of the core-collapse problem: the neutrino luminosity,
    $L_{\nu}$, mass accretion rate, $\dot{\mathcal{M}}$, neutron star
    radius or more specifically the neutrino-sphere radius, $R_{\nu}$,
    the neutrino temperature, $T_{\nu}$, and the neutron star mass,
    $M_{\rm NS}$.  Therefore, the explosion condition $\Psi_{\rm min}
    > 0$ represents a hypersurface in this five-dimensional parameter space.  By
    fixing 3 of the 5 parameters, one may construct ``critical
    curves'' with the other 2 parameters.  The critical $L_\nu-\dot{\mathcal{M}}$ (lower left panel) is one such
    example.  In each panel, the solid line shows the critical
    condition $\Psi_{\rm min} = 0$, and for all but the top panel
    explosions occur in the upper portion of the parameter space.  The
  dashed line shows the reduction of the critical condition due to
  neutrino-driven convection for each
  critical curve.
}
\label{triplot}
\epsscale{1.}
\end{figure*}
\epsscale{1.3}
\begin{figure}[t]
\hspace*{-.6cm}
\plotone{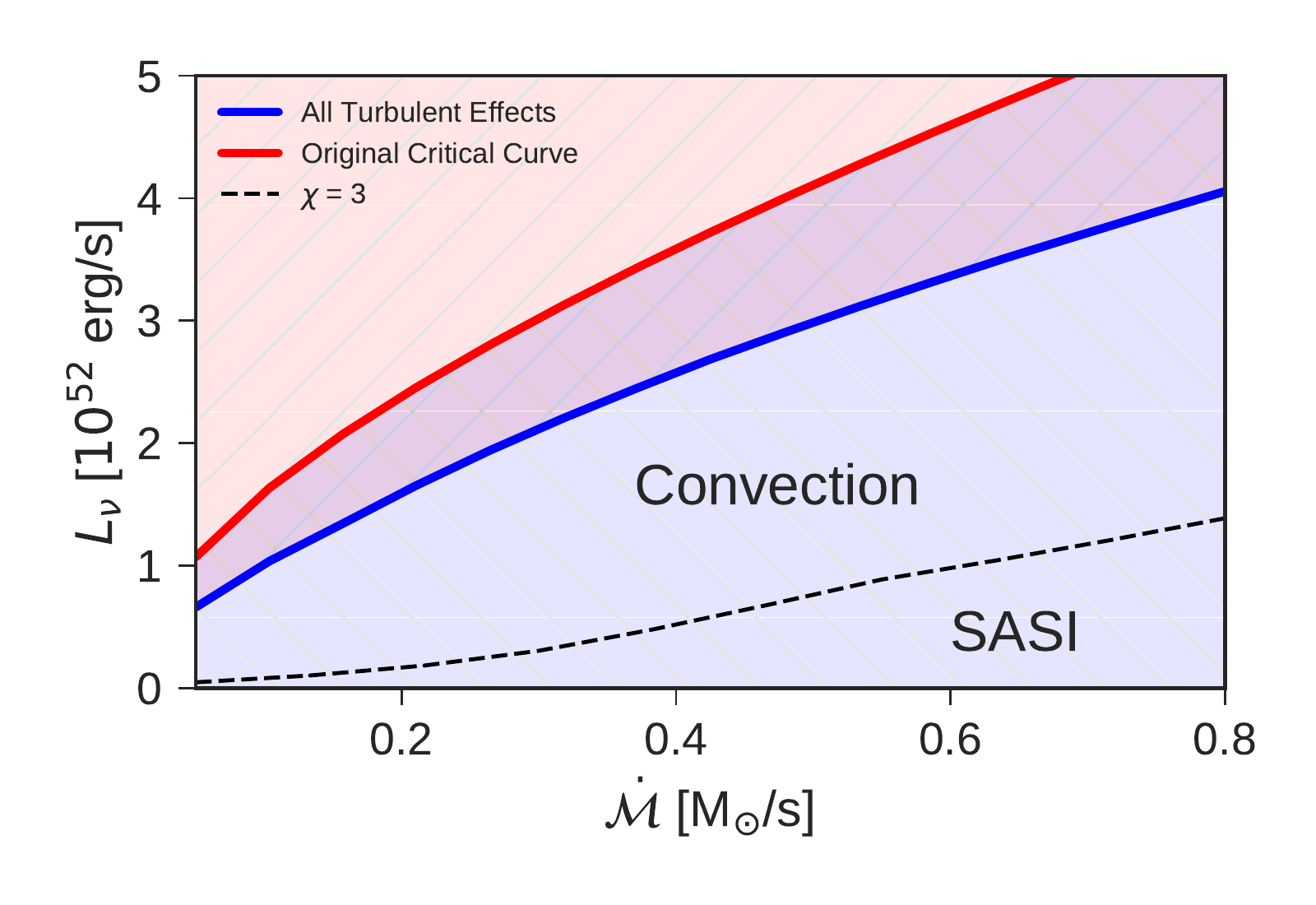}
\caption{Comparing the relative thresholds of the critical curves and
  the $\chi = 3$ line. Above this line, all stalled solutions have
  $\chi>3$ (convection dominated), and below this line, $\chi < 3$
  (SASI dominated).  The fact that convection dominates near the
  critical curve validates our use of a convection-based turbulence
  model to explore how turbulence affects the critical condition for
  explosions.}
\label{chi}
\end{figure}

\epsscale{.9}
\begin{figure*}[t]
\plotone{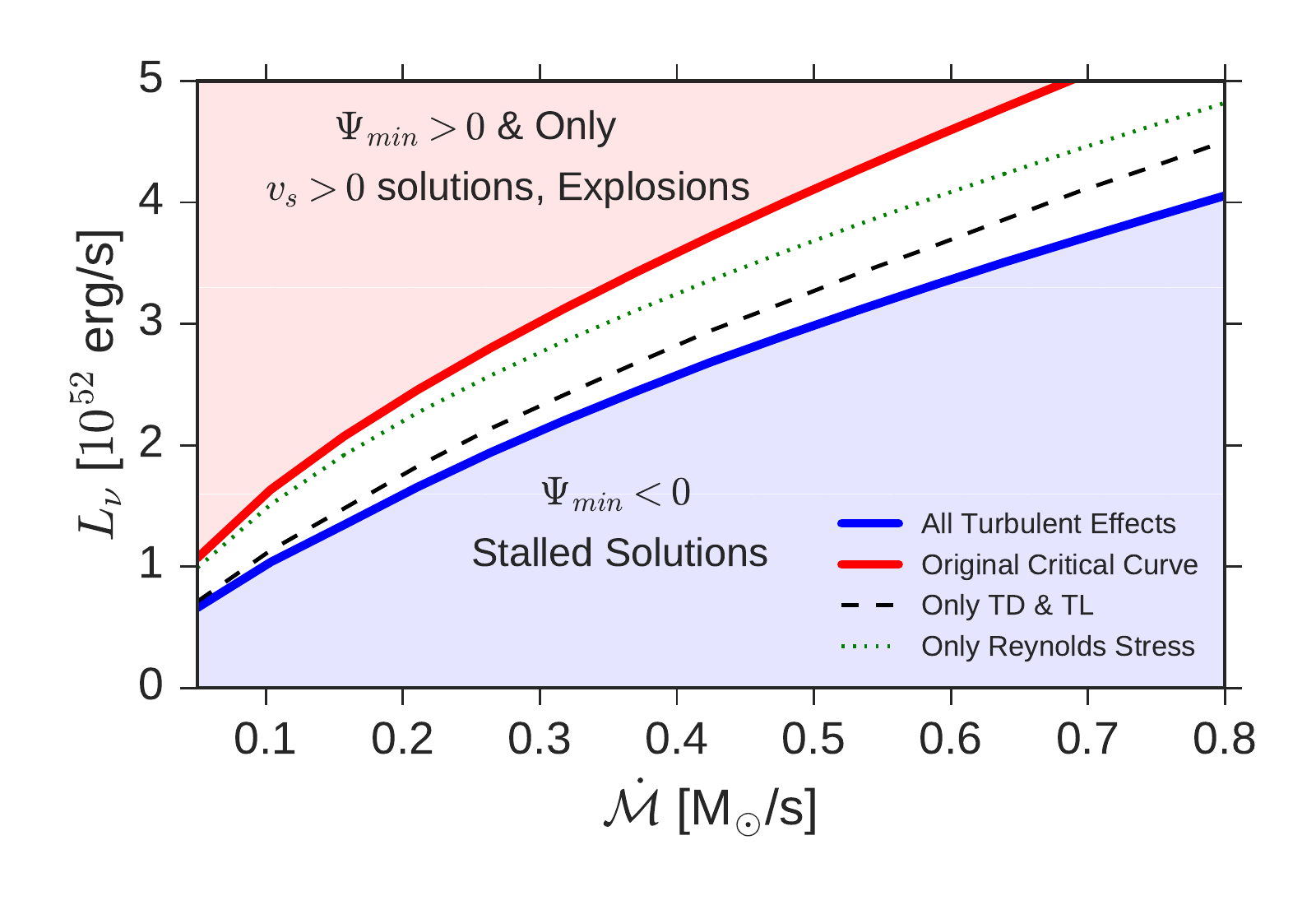}
\caption{Diagnosing how turbulence affects a
    critical curve for explosion.  Here we show
    how the various terms affect one particular slice of the
    $\Psi_{\rm min} = 0$ condition, the neutrino-luminosity
    vs. mass-accretion-rate critical curve.  The thick red
    line is the original critical curve \citep{burrows93}, and the thick
  blue line shows the turbulence induced reduction in the
    critical curve.  The
  red-shaded region is where $\Psi_{\rm min}>0$ and thus where no steady
  state solutions exist. The blue-shaded region is the region of
  stalled shock solutions when including convection.  To assess the effects of each of
    the turbulence terms, we reproduce the critical curve by isolating
  the turbulent terms in the energy and momentum equations.  In the
  energy equation, the terms are the turbulent dissipation
  and turbulent luminosity (TD and TL), and in the momentum equation,
  the only turbulent term is the Reynolds stress (or turbulent ram
  pressure). The dotted line
    corresponds to only including Reynolds stress, and the dashed line
    corresponds to only including the effects of TD and TL.  From
    these results we draw two main conclusions.  One, the necessary
    neutrino energy required for a supernova explosion is less when
    considering multi-dimensional effects.  Two, most of the reduction
    in the critical condition comes from the energy equation terms, in
  particular the turbulent dissipation.
\label{crit}}
\end{figure*} 
\epsscale{1.}

Our primary objective is to understand how turbulence affects
the conditions for explosion.  To fully understand this influence, we
also need to understand how turbulence affects the background
structure, so we first show how the turbulent terms affects the
density and temperature profiles. We then show that turbulence raises
the $\Psi$ parameter, implying an easing of the explosion
  condition. We then consider how this affects the critical
    hypersurface.  To
    connect to previous investigations, we focus on the
    neutrino-luminosity and accretion-rate slice of this critical
    hypersurface.  We find that this reduction in the critical curve is $\sim$30\%, in
    concordance with multi-dimensional simulations. To investigate how turbulence reduces the
    condition for explosion, we calculate the critical condition with each
    term included and omitted. Lastly, we provide evidence that our upper limit
on the Reynolds stress does not affect the actual reduction of the
critical curve. 

In Figure~\ref{normalized}, we illustrate how
turbulence affects the density profile;
in particular, we show the neutrino and convective terms in
  the derivative for the density.  Since the density profile most
  closely matches a power-law, we plot terms of d $\ln \rho/$d $\ln r$
  to compare how neutrinos and turbulence affect the slope in the
  log.  To reduce the clutter, we do not
  plot all of the terms; for the full ODE, see equation~(\ref{ODE1}).  In
  general, the missing terms give a slope that is about -3.  In the
  gain region, both neutrinos and the convective terms make the slope
  shallower.  In the cooling region, neutrino cooling makes the slope
  steeper. Turbulent dissipation
and luminosity terms ultimately originate from the energy
equation~(\ref{energy}).  The turbulent ram pressure term comes from the
momentum equation. While turbulent
  ram pressure does modify the density structure, the two turbulent terms from
  the energy equation, turbulent dissipation and turbulent luminosity, have a
  considerably larger effect on the structure.

Figure~\ref{denstemp} shows how turbulence affects the
  density and temperature profiles of the steady-state stalled shock
  solutions.  The net
effect of turbulence is to make the profiles much shallower.  In
part, turbulent ram pressure explains some of the difference,
but for the most part, turbulence provides extra heating through
  dissipation in the convective region.  One consequence is that the
  temperature (and entropy) are higher with turbulence.  This is
  consistent with the entropy profiles of multi-dimensional simulations
 \citep{murphy11,murphy13}. In agreement with simulations \citep{murphy13,abdi16}, Figure~\ref{denstemp} also shows that the solution including turbulence has a larger shock radius. The shock radius is a monotonic function of $L_\nu$, and since $\rho R_{rr}$ and
$\rho \epsilon$ effectively add energy in the same fashion as the
luminosity, we intuitively retrieve a larger shock radius.
Note that this larger shock radius is not just a consequence
  of turbulent ram pressure.  Instead, the post shock profile is
  shallower pushing out the shock.

Now that we understand how turbulence affects the density and
temperature profiles, we now present how turbulence affects the
critical curve in three figures. One, Figure~\ref{psi} shows how turbulence raises the dimensionless overpressure parameter, $\Psi$, in
\citet{murphy17}. Two, Figure~\ref{triplot}, shows
  how turbulence reduces the five-dimensional critical hypersurface for explosion
Finally, Figure~\ref{crit} shows that the dominant turbulent
  term in reducing the critical condition is turbulent dissipation. 

In Figure~\ref{psi} we plot $\Psi$ to show that the increase in the
  dimensionless parameter due to turbulence is roughly equivalent to
  increasing the neutrino luminosity by 30\%. To clarify, the $\Psi$ parameter is a measure of the
overpressure compared to the hydrostatic equilibrium. This
  integral condition is normalized by the
pre-shock ram pressure. This dimensionless
  parameter maps directly to the shock velocity in that when $\Psi$ is positive, the shock
expands, when $\Psi$ is negative, the shock stalls, and when $\Psi$ =
0, the shock has zero velocity. Note that there is always a
  minimum $\Psi$, and if this $\Psi_{\rm min}$ is greater than zero, then
  there are no stalled shock solutions, only steady expanding shocks.
 For the case where the minimum of
$\Psi$ is exactly zero, this set of solutions defines our critical
explosion condition for all parameters.  Clearly turbulence raises the minimum $\Psi$, and therefore would affect the critical curve. 

In figure~\ref{manycaps}, we show how the Reynolds stress upper limit
affects the explodability parameter and the critical curve. We suspected that, at high enough neutrino luminosities, the additional ram pressure at the shock would prevent finding solutions to the boundary conditions. Figure~\ref{manycaps} demonstrates that we consistently encounter this cap, but only for pseudo-solutions which have already found a critical $L_\nu$. We present several sets of solutions at different neutrino luminosities to emphasize that our $L_\nu-\dot{\mathcal{M}}$ critical curve is in fact not
affected by the upper limit on $\mathcal{R}$, even at unrealistic luminosities. The sole determination of the critical curve is on
$\Psi_{\rm min} = \Psi(r_{\rm s}^{\rm crit})$, and since this threshold is only reached when $r_{\rm s} > r_{\rm s}^{\rm crit}$, there is no effect on the critical curve. 

Figure~\ref{triplot} shows how turbulence
modifies the critical hypersurface. \citet{murphy17} points out that
  the critical condition for explosion is not a critical curve, but a
  critical hypersurface in a five-dimensional space that is defined by
  one dimensionless condition: $\Psi_{\rm min} = 0$.  The neutrino-luminosity-accretion-rate curve
is one slice of this critical hypersurface.  In Fig.~\ref{triplot}, we
show six slices of this hypersurface.  Note that in all panels except the
  top-left panel ($M_{\rm NS}$ vs. $\mdot$), the region of explosion
  is above the curve. For $M_{\rm NS}$ vs. $\mdot$, it is below the
curve; a lower mass neutron star has a lower potential to overcome to
explode. Because turbulence raises the
  $\Psi$ curves, it also reduces the critical hypersurface for
  explosion.
  
In figure \ref{chi}, we make a case for the validity of using a
convection-based turbulence model. The $\chi$ parameter, first
introduced by \citet{fog06}, is a measure of the linear stability of
the convective region in the presence of advection.  For $\chi < 3$,
advection stablizes the flow and convection does not mainfest.  The
assumption is that since convection is suppressed, the SASI dominates
the turublence.  The $\chi$ parameter is defined as
\begin{equation}
  \chi \equiv \int_{r_g}^{r_s} \left | \frac{\omega_{buoy}}{u} \right
  | \, dr
\end{equation}
where $\omega_{buoy}$ is the Brunt-V\"ais\"al\"a frequency, defined as 
\begin{equation}
  \omega^2_{buoy} \equiv \left [ \frac{1}{p}\left ( \frac{\partial p}{\partial S}\right )_{\rho,Y_e} \frac{dS}{dr} + \frac{1}{p}\left ( \frac{\partial p}{\partial Y_e}\right )_{\rho,S}  \frac{dY_e}{dr} \right ] \frac{g}{\Gamma_1} \,   
\end{equation}
and
\begin{equation}
\Gamma_1 \equiv \left ( \frac{\partial \ln P}{\partial \ln \rho} \right )_{S,Y_e} \, .
\end{equation} 
In Figure~\ref{chi}, above the $\chi = 3$ dashed line, all
stalled solutions have $\chi > 3$
(convection dominated) and below this dashed line, all stalled
solutions have $\chi < 3$ (SASI
dominated).  To calculate this line, we use a similar approach as deriving the critical luminosity curves: we input all
of our parameters, and numerically solve for the luminosity which
gives a value of $\chi = 3$.  According to these
models, convection dominates for all scenarios near explosion.  Recent
simulations seem to support this conclusion
\citep{abdi15,lentz15,couch14,burrows12,iwakami14,ott13,roberts16}. The
results of Figure~\ref{chi} validate our exploration of how
convection-based turbulence affects the critical condition.  However,
we do encourage future explorations of the SASI and $\chi$ in
simulations to validate wether convection is indeed dominant near the
critical condition.

In figures \ref{psi} and \ref{triplot}, we considered the overall
effect of turbulence on the critical condition, but this does
not illuminate how turbulence reduces the condition for explosion. In
figure~\ref{crit}, we focus exclusively on the
$L_\nu-\dot{\mathcal{M}}$ critical curve and explore the
  effect of each term in this slice of the critical condition.  We
find that turbulent dissipation within the gain region acts as an
even greater driving force for explosion than the turbulent ram
pressure. That is not to say that the Reynolds stress is negligible;
both terms are indeed needed to obtain the critical curve reduction
predicted by multi-dimensional simulations. Though this result relies
upon some assumptions in the turbulence model, we suspect that the qualitative outcome
  will persist: turbulent dissipation can not be dismissed.  


\subsection{Buoyancy Driven Heating}


We argue that turbulent dissipation adds another heat source
  which aids explosion.  Since neutrinos are the ultimate source of
  energy, it may seem that we are asking neutrinos to do twice the
  work.  However, this is not the case. Using a simple convective
  model, we propose that in one dimension, some of the neutrino energy goes into
  heating the material, and some of it goes into creating a higher
  potential profile.  In multiple dimensions, this higher
    potential profile is unstable, goes to a lower potential profile,
    and the excess energy goes into kinetic energy, which in turn
    dissipates as heat via turbulent dissipation.

\begin{figure*}[t]
\plotone{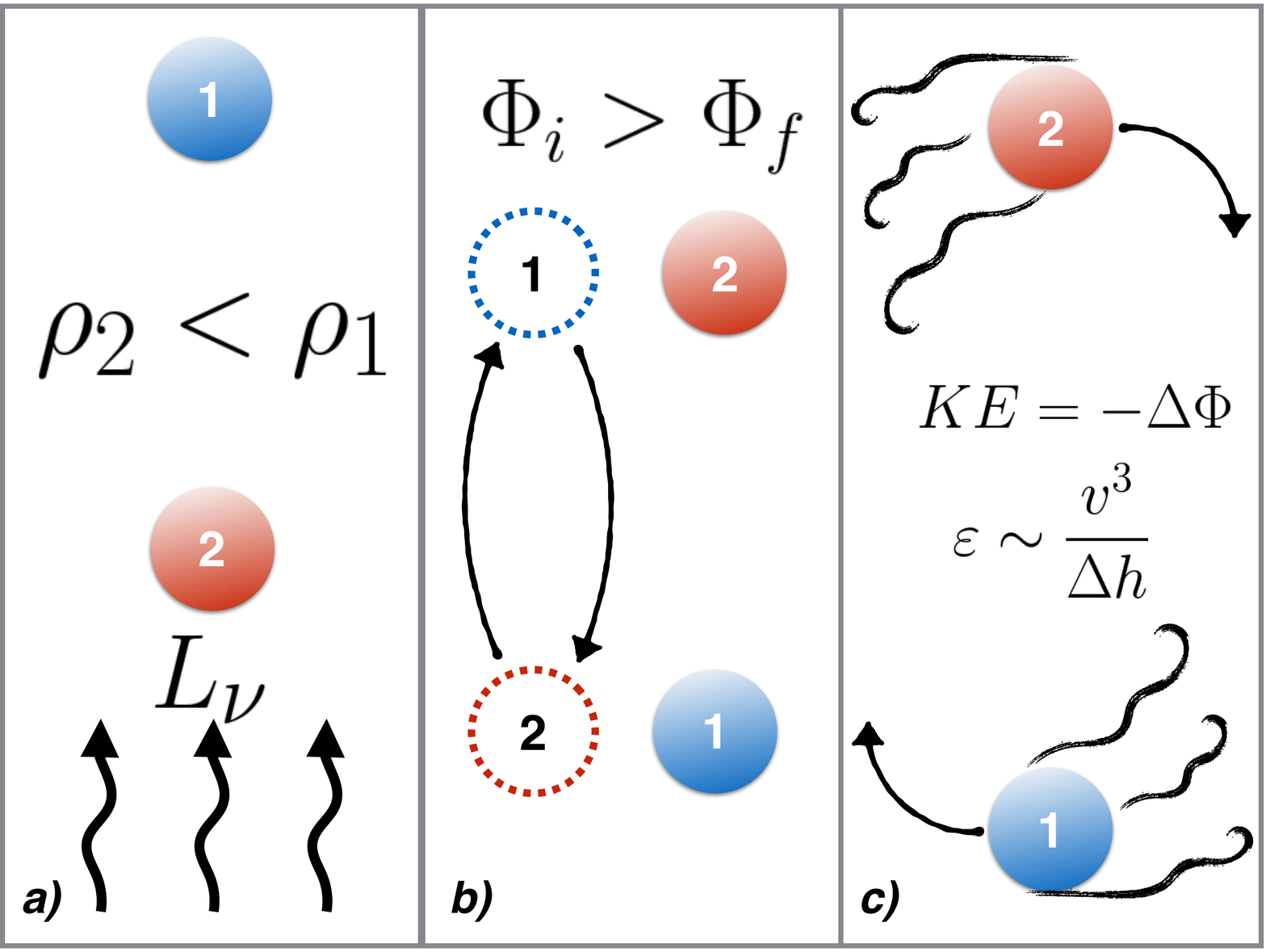}
\caption{Neutrinos heat the gain region and setup a higher potential
  state, which turbulence taps and dissipates as heat.
    Panel a): when a parcel of matter advects through the gain region,
    neutrinos heat it, which sets up a buoyantly unstable situation.  Panel b): Parcel 1 wants to buoyantly sink and parcel 2
  wants to buoyantly rise.  The final state has a lower potential
  energy than the final state.  Panel c): This change in potential
  energy is converted to turbulent kinetic energy which dissipates via
  turbulent dissipation.  Therefore, in 1D, part of the energy of
  neutrinos heats the gain region and part of the neutrino energy goes
  into setting up the higher potential profile.  Multi-dimensional
  turbulence taps into this higher potential energy by allowing for a
  lower potential state and turbulent kinetic energy.  Then that
  turbulent kinetic energy is dissipated through viscosity. }
\label{RBC}
\end{figure*}

Rayleigh-B\'enard convection is a simple convective model which can
clearly demonstrate this conversion from potential to kinetic to
dissipated internal energy. Figure~\ref{RBC} illustrates the
fundamental physics of convection. First, neutrinos provide a source
of heating and drives a convective instability. In this cartoon model,
we consider two parcels; the lower one receives more neutrino heating,
has a higher entropy, and lower density compared to its
surroundings. The parcel at a higher height has a lower entropy and higher density
compared to its surroundings. If one switches the positions of these
two parcels, then one finds that the gravitational potential is
lower. Therefore, neutrino heating causes a higher potential structure
that is unstable to convective overturn. The difference in potentials
between the two states gets converted into kinetic energy of the
parcels. Kolmogorov's hypothesis suggests that the dissipation of this
kinetic energy is $R^{3/2}/\mathcal{L}$ and happens on the order of
one turnover timescale.  \citet{burrows95} also considered this idea in
which two layers of varying densities are swapped, inducing a buoyant
work being done on the system. This energy is then converted into
kinetic energy in the form of eddies, and in turn dissipated into
heat. Therefore, not only do neutrinos heat the gain region, but they
also create a higher potential system. This higher potential gets
converted to kinetic energy and consequently dissipated as internal
energy.

To illustrate this more clearly, consider the energy equation (\ref{energy}).  It is
more illuminating if we rewrite the equation considering a constant
mass accretion rate:
\begin{equation}
\dot{\mathcal{M}}\partial_i \left [h + \frac{u^j u_j}{2} - \Phi \right ] =  \rho q
\end{equation} 
Neutrinos heat the convective region of the star, changing the enthalpy and gravitational potential. This gives rise to an entropy and density gradient such that $S_2 > S_1$ and $\rho_2 < \rho_1$ (thus satisfying the Schwarzschild condition for convection). We then treat the potential energy of two parcels in a similar manner to \citet{burrows95}. Before the exchange of parcels, the gravitational potential energy is
\begin{equation}
  \Phi_1 = gV(\rho_1 h_1 + \rho_2 h_2)
\end{equation}
But after the top parcel sinks and the bottom parcel buoyantly rises, the potential is then
\begin{equation}
  \Phi_2 = gV(\rho_2 h_1 + \rho_1 h_2)
\end{equation}
Thus, the change in potential is
\begin{equation}
  \Delta \Phi = gV(\rho_2 - \rho_1)(h_1 - h_2)
\end{equation}
Since $h_1 > h_2$ and $\rho_1 > \rho_2$, $\Delta \Phi < 0$.
Conservation of energy suggests that K.E.$\approx$-$\Delta \Phi$, and via turbulent dissipation, this
energy from buoyant driving is converted into heat. In summary, neutrinos heat the gain region and setup a higher potential profile. Turbulence allows a lower potential state and the kinetic energy is converted into thermal energy via turbulent dissipation, aiding explosion.

\section{Conclusion}
\label{section4}

     

A major result of core-collapse theory is
that one-dimensional simulations fizzle for all but the least
massive stars, while multi-dimensional
  simulations explode for the most part.
  Even though there are still some differences between the
    simulations, there is a general consensus that turbulence makes the difference between a fizzled outcome and a successful
    explosion. To explain how turbulence aids explosion, we
  develop a turbulence model for neutrino-driven convection \citep{murphy13}
  and investigate how turbulence reduces the critical condition for
  explosion \citep{burrows93,murphy17}.   Our turbulence model
  reduces the critical condition for explosion by $\sim$30\%, in general
  agreement with simulations. By modeling each turbulent
    term, we are able to investigate the effect of each turbulent term
  on the critical condition. We find that although ram
  pressure plays a role in reducing the critical curve, it is not the
  dominant term.  The dominant term in reducing the critical curve is
  turbulent dissipation. Furthermore, we are not the first to suggest that turbulent dissipation is important in aiding neutrino-driven explosions.  \citet{thompson05} suggested that MRI driven turbulence for very rapidly rotating proto-neutron stars could add significant heating and aid explosion.

In the turbulence model, we include all turbulent terms, both in the
  background solution and in the boundary conditions.
The three main turbulent terms are turbulent dissipation, Reynolds
stress, and turbulent luminosity. Overall, we find that all three play an important role in modifying the background
  structure and the critical condition for explosion.  However, it is
  the terms in the energy equation, the combined turbulent luminosity and turbulent dissipation, which give the largest reduction in the
  critical curve. Of these two, the turbulent dissipation provides the largest effect
  in reducing the critical condition. 
  
The ultimate source of power for convection and turbulent dissipation
is the neutrino driving power.  This may seem as if we are double
counting the power supplied by neutrinos.  However, we are not.
Instead, we propose that neutrinos heat the gain region and setup a
higher potential, convectively unstable structure.  In
multi-dimensional simulations, convection converts this higher
potential structure to a lower potential structure.  The change in
potential energy is converted to kinetic turbulent energy, which in
turn dissipates as heat.  We suggest that CCSN modelers check the
structures of their one-dimensional and multi-dimensional simulations
to test this supposition.

The turbulence model is a global, integral model, and provides
  little constraint on the local structure of turbulence, but
    a local model is necessary in solving the background
    equations and deriving a critical condition.  Therefore,  
we made some fairly straightforward assumptions to translate the
  global model to a local model. For example, we
    assumed that the specific turbulent dissipation rate is
      constant throughout the gain region. We also had to assume a
      specific spatial profile for the turbulent luminosity.  These assumptions probably do
    not affect our qualitative results.  However, our results have large implications regarding how turbulence  aids explosion.  In particular, we propose that turbulent
  dissipation is a key contributor to reviving a stalled shock to a
  successful supernova.  Therefore, these assumptions
    shoud be verified with multi-dimensional simulations and treated more rigorously in future
    investigations.

To verify the predictions of this manuscript, we identify at least three open
  questions that multi-dimensional simulations should address.  One, does turbulent
  dissipation actually lead to significant heating?
In multiple dimensions, the entropy profile should be a result of neutrino heating and cooling, turbulent entropy flux, and turbulent
    dissipation.  One can easily calculate the turbulent entropy flux,
    and the heating and cooling by neutrinos.  What ever is left
    should be equal to the expected entropy generated by turbulent
    dissipation. Second, do the one-dimensional profiles
      have a higher potential compared to their multi-dimensional
      counterparts? Third, do the local
      details of the turbulent model matter?  We suspect that the
    local details do not change the
    qualitative result. However, one should compare our local assumptions with multi-dimensional
      simulations, and assess how (if at all) the
    quantitative results vary from this work.

In summary, combining a turbulence model and critical
  condition analyses helps to illuminate how turbulence aids
  explosions.  Specifically, by modeling each turbulent term we are
  able to assess the effect of each term in the conditions for
  explosion.  Contrary to prior suppositions, we find that turbulent
  ram pressure is not the dominant effect.  Rather, each of the three
  terms are quite large in their effect with turbulent dissipation
  being the largest.  Presently, these conclusions are qualitative.
  To be predictive, the community will need to verify these
  conclusions with multi-dimensional simulations.  Eventually, we may
  be able to use these turbulence models to make one-dimensional
  simulations explode under similar conditions as multi-dimensional
  simulations, thereby enabling rapid and systematic exploration of
  the explosion of massive stars.

\section*{Acknowledgments}
We would like to thank Joshua C. Dolence for illuminating discussions
on the effect of turbulence on the critical condition.  We are also
thankful to David Radice for challenging us to explain the source of
energy for turbulent dissipation.  This material is based upon work supported by the National Science Foundation under Grant No. 1313036.

\appendix

\section{Full ODEs}
\label{sec:ODEs}
Here, we present the full set of ordinary differential equations that
we solve to find pseudo steady-state solutions.  The equation for
density is
\begin{equation}
\label{ODE1}
      \scalebox{1.3}{$\frac{dln \rho}{dlnr} = \frac{\frac{2v^2}{\Phi} - 1 +
    \mathcal{RS} + \frac{y}{l}\frac{\frac{\partial lny}{\partial
        lnT}}{\frac{\partial lnl}{\partial lnT}}\big(1 + \mathcal{H} -
    \mathcal{C} + \mathcal{TD} - \mathcal{TL} +
    \frac{2v^2}{\Phi}\big)}{y - \frac{v^2}{\Phi} + y \frac{\partial
      lny}{\partial ln \rho} + \frac{\frac{\partial lny}{\partial
        lnT}}{\frac{\partial lnl}{\partial lnT}}\big(\frac{y v^2}{\Phi
     l}  - \frac{\partial lnl}{\partial ln \rho}\big)}$} \, ,
\end{equation}
and in terms of the density equation, the temperature ODE is
\begin{equation}
\label{ODE2}
\scalebox{1.3}{$\frac{\partial \ln T}{\partial \ln r} = \frac{\big(1 +
    \mathcal{H} - \mathcal{C} + \mathcal{TD} - \mathcal{TL} +
    \frac{2v^2}{\Phi}\big)}{l \frac{\partial lnl}{\partial lnT}} +
  \frac{\frac{\partial ln \rho}{\partial lnr}}{\frac{\partial
      lnl}{\partial lnT}} \bigg( \frac{v^2}{\Phi l} - \frac{\partial
    lnl}{\partial ln \rho} \bigg)$} \, .
\end{equation}

In this form, these equations seem some what unwieldy, but
  they would be even more so if we had not made the following
  shorthand for the important physics.  To further help illustrate the
important scales in the problem, we present each important physics in
terms of a dimensionless variable.
The Reynolds stress (or ram pressure) appears in the above
  equations as
\begin{equation}
\mathcal{RS} = \frac{2x^2R_{rr}}{(x_s-x_g)\Phi_1} \, .
\end{equation}
In these expressions, the subscript 1 indicates the base of the
solution.  In our particular case, that is the neutrino-sphere
radius.  Neutrino heating and cooling are
\begin{equation}
\mathcal{H} = \frac{L_\nu \kappa \rho r_1 x^2}{\dot{\mathcal{M}}
  \Phi_1}  \, ,
\end{equation}
and
\begin{equation}
\mathcal{C} = \frac{C_0 \big( \frac{T}{T_0} \big )^6 4 \pi r_1^3 x^4
  \rho}{\dot{\mathcal{M}} \Phi_1} \, .
\end{equation}
The two turbulent terms from the energy equation are the
  turbulent dissipation and turbulent luminosity:
\begin{equation}
\mathcal{TD} = \frac{L_e^{max} r_1^3 \rho w_b 4 \pi}{M_g \dot{\mathcal{M}} \Phi_1}
\end{equation}
and
\begin{equation}
\mathcal{TL} = \frac{L_e^{max} r_1 x^2}{cosh ^2 \big( \frac{(r_1(x -
    x_g))}{h} \big) h \dot{\mathcal{M}} \Phi_1} \, .
\end{equation}
Two dimensionless measures of the pressure and enthalpy are
\begin{equation}
  y = \frac{P}{\rho \Phi}
\end{equation}
and
\begin{equation}
  l = \frac{e + \frac{P}{\rho}}{\Phi}
\end{equation}
\begin{equation}
  P_{,\rho} = \frac{\partial \ln P}{\partial \ln \rho} \bigg |_T
\end{equation}
The normalized radius is
\begin{equation}
  x = \frac{r}{r_{NS}} \, .
\end{equation}
A dimensionless measure of the bouyant driving is
\begin{equation}
  w_b = \int\limits_{x_g}^{x_s} \frac{tanh \big( \frac{r_1(x -
      x_g)}{h} \big)}{3yx} dx \, .
  \label{endofeq}
\end{equation}
To observe and distinguish the effects of each turbulent
term, we add the capability to turn
  each term on or off in the equations. In doing so, we investigate the effect
of each term on the solutions and critical curves.
\\

\end{document}